\documentclass{iopart}
\usepackage{iopams,setstack}
\usepackage{graphicx}

\newcommand{\LQCD}{\Lambda_{\rm QCD}}
\newcommand{\Nc}{N_{\rm c}}
\newcommand{\Nf}{N_{\rm f}}
\newcommand{\Tc}{T_{\rm c}}
\newcommand{\Td}{T_{\rm d}}

\newcommand{\Tpc}{T_{\rm pc}}
\newcommand{\TE}{T_{\rm E}}
\newcommand{\TF}{T_{\rm F}}
\newcommand{\TG}{T_{\rm G}}
\renewcommand{\TH}{T_{\rm H}}
\newcommand{\mc}{m_{\rm c}}
\newcommand{\muc}{\mu_{\rm c}}
\newcommand{\mue}{\mu_{\rm e}}
\newcommand{\muE}{\mu_{\rm E}}
\newcommand{\muF}{\mu_{\rm F}}
\newcommand{\muG}{\mu_{\rm G}}
\newcommand{\muH}{\mu_{\rm H}}
\newcommand{\muI}{\mu_{\rm I}}

\newcommand{\Dud}{\Delta_{ud}}
\newcommand{\Dds}{\Delta_{ds}}
\newcommand{\Dsu}{\Delta_{su}}

\newcommand{\nB}{n_{\rm B}}
\newcommand{\nq}{n_{\rm q}}
\newcommand{\Nq}{N_{\rm q}}

\newcommand{\muB}{\mu_{\rm B}}
\newcommand{\muq}{\mu_{\rm q}}
\newcommand{\muqim}{\tilde{\mu}_{\rm q}}

\newcommand{\mq}{m_{\rm q}}
\newcommand{\Mq}{M_{\rm q}}
\newcommand{\MB}{M_{\rm B}}

\newcommand{\mst}{m_{\rm s}}
\newcommand{\mup}{m_{\rm u}}
\newcommand{\mdw}{m_{\rm d}}
\newcommand{\mud}{m_{\rm ud}}
\newcommand{\fq}{f_{\rm q}}
\newcommand{\fqb}{f_{\bar{\rm q}}}
\newcommand{\fqq}{f_{\bar{\rm q}\rm q}}

\newcommand{\diag}{\mathrm{diag}}

\newcommand{\MeV}{\,\mbox{MeV}}
\newcommand{\GeV}{\,\mbox{GeV}}
\newcommand{\fm}{\,\mbox{fm}}

\newcommand{\bx}{\boldsymbol{x}}
\newcommand{\by}{\boldsymbol{y}}
\newcommand{\bp}{\boldsymbol{p}}
\newcommand{\bq}{\boldsymbol{q}}

\newcommand{\bxs}{\mbox{\scriptsize \boldmath $x$}}
\newcommand{\bys}{\mbox{\scriptsize \boldmath $y$}}

\newcommand{\ua}{\mathrm{U}(1)_{\rm A}}
\newcommand{\one}{1\kern-0.36em1}
\newcommand{\calG}{\mathcal{G}}
\newcommand{\calH}{\mathcal{H}}

\newcommand{\psiR}{\psi_{\rm R}^{\phantom{\dagger}}}
\newcommand{\psiL}{\psi_{\rm L}^{\phantom{\dagger}}}
\newcommand{\dL}{d_{\rm L}^{\phantom{\dagger}}}
\newcommand{\dR}{d_{\rm R}^{\phantom{\dagger}}}
\newcommand{\dLd}{d_{\rm L}^\dagger}
\newcommand{\dRd}{d_{\rm R}^\dagger}

\newcommand{\Qtilde}{\tilde{Q}}

\newcommand {\apgt} {\ {\raise-.5ex\hbox{$\buildrel>\over\sim$}}\ } 
\newcommand {\aplt} {\ {\raise-.5ex\hbox{$\buildrel<\over\sim$}}\ }

\begin{document}


\begin{flushright}
\textsf{YITP-10-28, TKYNT-10-06}
\end{flushright}

\title[The phase diagram of dense QCD]
      {The phase diagram of dense QCD}

\author{Kenji Fukushima$^{1}$ and Tetsuo Hatsuda$^{2}$}

\address{$^1$ Yukawa Institute for Theoretical Physics,
         Kyoto University, Oiwake-cho, Kitashirakawa,
         Sakyo-ku, Kyoto-shi, Kyoto 606-8502, Japan}
\address{$^2$ Department of Physics,
         The University of Tokyo, 7-3-1 Hongo,
         Bunkyo-ku, Tokyo 113-0033, Japan}
\ead{fuku@yukawa.kyoto-u.ac.jp}
\ead{hatsuda@phys.s.u-tokyo.ac.jp}

\begin{abstract}
 Current status of theoretical researches on the QCD phase diagram at
 finite temperature and baryon chemical potential is reviewed with
 special emphasis on the origin of various phases and their symmetry
 breaking patterns.  Topics include;  quark deconfinement, chiral
 symmetry restoration, order of the phase transitions, QCD critical
 point(s), colour superconductivity, various inhomogeneous states and
 implications from QCD-like theories.
\end{abstract}

\submitto{\RPP}
\maketitle

\tableofcontents
\markboth{The phase diagram of dense QCD}{The phase diagram of dense QCD}


\section{Introduction}

One of the most crucial properties in the non-Abelian gauge theory of
quarks and gluons, Quantum Chromodynamics (QCD) \cite{Nambu:1981pg},
is the asymptotic freedom \cite{Politzer:1973fx,Gross:1973id}; the
coupling constant runs towards a smaller value with increasing energy
scale.  It is hence a natural anticipation that QCD matter at high
energy densities undergoes a phase transition from a state with
confined hadrons into a new state of matter with on-shell (real)
quarks and gluons.

There are two important external parameters for QCD in equilibrium,
the temperature $T$ and the baryon number density $\nB$.  (In the
grand canonical ensemble, the quark chemical potential $\muq=\muB/3$
may be introduced as a conjugate variable to the quark number density
$\nq= 3\nB$.)  Since the intrinsic scale of QCD is $\LQCD\sim200\MeV$,
it would be conceivable that the QCD phase transition should take
place around $T\sim\LQCD\sim\mathcal{O}(10^{12})\,{\rm K}$ or
$\nB\sim\LQCD^3\sim1\fm^{-3}$.  Experimentally, the heavy-ion
collisions (HIC) in laboratories provide us with a chance to create
hot and/or denser QCD matter and elucidate its properties.  In
particular, the Relativistic Heavy-Ion Collider (RHIC) at Brookhaven
National Laboratory (BNL) has conducted experiments to create hot QCD
matter (a quark-gluon plasma or QGP in short) by the Au-Au collisions
with the highest collision energy $\sqrt{s_{_{NN}}}=200\GeV$
\cite{Stankus:2009zz}.  The Large Hadron Collider (LHC) at CERN will
continue experiments along the same line with higher energies
\cite{Stankus:2009zz,Abreu:2007kv}.  Exploration of a wider range of
the QCD phase diagram with $\nB$ up to several times of the normal
nuclear matter density $n_0\simeq0.17\fm^{-3}$ may be carried out by
low-energy scan in HIC at RHIC as well as at the future facilities
such as the Facility for Antiproton and Ion Research (FAIR) at GSI,
the Nuclotron-based Ion Collider Facility (NICA) at JINR and the Japan
Proton Accelerator Research Complex (J-PARC) at JAERI.

In nature, the deep interior of compact stellar objects such as
neutron stars would be the relevant place where dense QCD matter at
low temperature is realized (see \cite{Heiselberg:1999mq} for a
review).  In fact, continuous efforts have been and are being
conducted in the observations of neutron stars to extract information
of the equation of state of dense QCD.\ \ If the baryon density is
asymptotically high, weak coupling QCD analyses indicate that the QCD
ground state forms a condensation of quark Cooper pairs, namely the
colour superconductivity (CSC).  Since quarks have not only spin but
also colour and flavour quantum numbers, the quark pairing pattern is
much more intricate than the electron pairing in metallic
superconductors.

In this review we will discuss selected topical developments in the
QCD phase diagram with an emphasis on the phases at finite $\muB$.
For the topics not covered in the present article, readers may consult
other reviews
\cite{Gross:1980br,Svetitsky:1985ye,Klevansky:1992qe,Hatsuda:1994pi,%
MeyerOrtmanns:1996ea,Smilga:1996cm,Rajagopal:2000wf,Rischke:2003mt,
Alford:2007xm,Hayano:2008vn,Satz:2009hr,Huang:2010nn,Schmitt:2010pn}.

We organize this article as follows.  In \sref{sec:overview}, after a
brief introduction of two key features of hot/dense QCD, i.e.\ the
deconfinement and chiral restoration, we show a conjectured QCD phase
diagram in $\muB$--$T$ plane.  In \sref{sec:confinement} we introduce
three major order parameters to characterize QCD matter at finite $T$
and $\muB$, i.e.\ the Polyakov loop, the chiral condensate and the
diquark condensate.  In \sref{sec:uncertain} we summarize the current
status on the chiral phase transition at finite $T$ with $\muB=0$.  In
\sref{sec:chiral} we go into more details about the phase transitions
in the density region up to a moderate $\muB$ comparable to $T$, which
are classified according to different spontaneous chiral symmetry
breaking (S$\chi$B) patterns.  We then address the situation with
$\muB$ much greater than $T$ in \sref{sec:diquark}.  Even though the
theoretical understanding has not been fully settled down yet,
\sref{sec:inhomogeneous} is devoted to a review over conjectured
inhomogeneous states of QCD matter as well as some alternative
scenarios.  In \sref{sec:suggestions} we look over suggestive results
and implications from QCD-like theories.  \Sref{sec:summary} is
devoted to the summary and concluding remarks.


\section{QCD phase structure}
\label{sec:overview}

We will run through the QCD phase transitions and associated phase
structure here before detailed discussions in subsequent sections.


\subsection{Deconfinement and chiral restoration}
\label{sec:overview1}

A first prototype of the QCD phase diagram in $T$-$\nB$ plane was
conjectured in \cite{Cabibbo:1975ig}.  It was elucidated that one
could give an interpretation of the Hagedorn's limiting temperature in
the Statistical Bootstrap Model (SBM) \cite{Hagedorn:1965st} as a 
critical temperature associated with a second-order phase transition
into a new state of matter.  The weakly interacting quark matter at
large $\nB$ due to the asymptotic freedom had been also recognized
\cite{Collins:1974ky}.  Historical summary of QCD phase diagram and
its exploration in HIC experiments are given in the reviews
\cite{Baym:2001in,Fukushima:2008pe}.

\vspace{.5em}

\begin{itemize}

\item
\textit{Deconfinement} ---
In an early picture of hadron resonance gas at finite temperature
\cite{Hagedorn:1984hz}, the density of (mostly mesonic) states
$\rho(m)$ as a function of the resonance mass $m$ is proportional to
$\exp(m/T^{\rm H})$ where $T^{\rm H}\simeq0.19\GeV$ is known from the
Regge slope parameter.  Such an exponentially growing behaviour of the
density of states should be balanced by the Boltzmann factor
$\exp(-m/T)$ in the partition function.  When $T>T^{\rm H}$, the
integration over $m$ becomes singular, so that $T^{\rm H}$ plays a
role of the limiting temperature (Hagedorn temperature) above which
the hadronic description breaks down.  This argument is applied to
estimate the critical value of $\muB$ as well.  The density of
baryonic states,
$\rho_{\rm B}(m_{\rm B})\propto \exp(m_{\rm B}/T_{\rm B}^{\rm H})$, is
balanced by the Boltzmann factor $\exp[-(m_{\rm B}-\muB)/T]$, leading
to the limiting temperature $T=(1-\muB/m_{\rm B})T_{\rm B}^{\rm H}$.
We see equivalently that the critical $\muB$ at $T=0$ is given by
$m_{\rm B} (\apgt 1\GeV)$.

If one bears in mind a simple bag-model picture that hadrons are
finite-size objects in which valence quarks are confined, it is
conceivable to imagine that hadrons overlap with each other and start
to percolate at the Hagedorn temperature \cite{Baym:1979,Satz:1998kg},
which is an intuitive portrayal of quark deconfinement.  Although the
Hagedorn/percolation picture is useful for practical objectivization,
it is necessary to develop a field-theoretical definition of the
quark-deconfinement in QCD.\ \ Global centre symmetry of pure gluonic
sector of QCD gives such a definition as elucidated in
\sref{sec:pol-dec}.

\vspace{.5em}

\item
\textit{Chiral restoration} ---
The QCD vacuum should be regarded as a medium with full of quantum
fluctuations that are responsible for the generation of
non-perturbative quark mass.  In hot and dense energetic matter,
quarks turn bare due to asymptotic freedom.  Therefore, one may expect
a phase transition from a state with heavy \textit{constituent} quarks
to another state with light \textit{current} quarks.  Such a
transition is called chiral phase transition named after the
underlying chiral symmetry of QCD.\ \ The QCD phase diagram at finite
$T$ and $\muB$ was also conjectured from the point of view of chiral
symmetry \cite{Hatsuda:1985eb}.  In this case, the order parameter is
the chiral condensate $\langle\bar{\psi}\psi\rangle$ which takes a
value about $-(0.24\GeV)^3$ in the vacuum and sets a natural scale for
the critical temperature of chiral restoration.  In the chiral
perturbation theory ($\chi$PT) the chiral condensate for two massless
quark flavours at low temperature is known to behave as
$\langle\bar{\psi}\psi\rangle_T/\langle\bar{\psi}\psi\rangle
 =1-T^2/(8f_\pi^2)-T^4/(384f_\pi^4)-\cdots$ with the pion decay
constant $f_\pi\simeq93\MeV$ \cite{Gerber:1988tt}.  Although the
validity of $\chi$PT is limited to low temperature, this is a clear
evidence of the melting of chiral condensate at finite temperature.
At low baryon density, likewise, the chiral condensate decreases as
$\langle\bar{\psi}\psi\rangle_{\nB}/\langle\bar{\psi}\psi\rangle
 = 1-\sigma_{\pi N}\,\nB/(f_\pi^2 m_\pi^2)-\cdots$
\cite{Drukarev:1991fs,Cohen:1991nk,Hatsuda:1991ez} where
$\sigma_{\pi N}\sim40\MeV$ is the $\pi$-$N$ sigma term.  (For
higher-order corrections, see \cite{Kaiser:2007nv,Kaiser:2008qu}.)

The chiral transition is a notion independent of the deconfinement
transition.  In \sref{sec:chiral-breaking} we  classify the chiral
transition according to the S$\chi$B pattern.

\vspace{.5em}

\end{itemize}


\subsection{Conjectured QCD phase diagram}
\label{sec:overview2}


\begin{figure}
 \begin{center}
\includegraphics[width=0.85\textwidth]{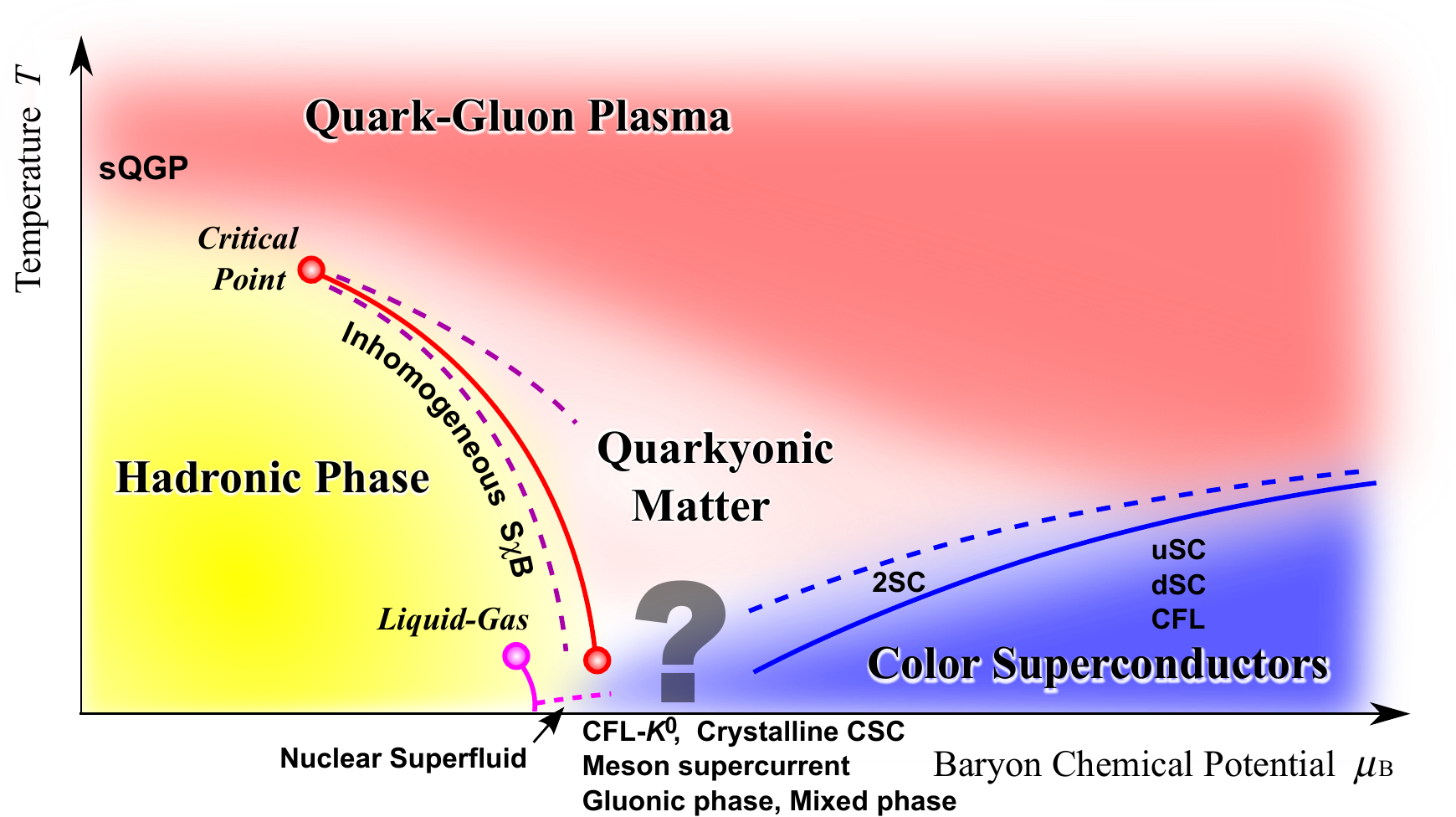}
 \end{center}
 \caption{Conjectured QCD phase diagram with boundaries that define
   various states of QCD matter based on S$\chi$B patterns.}
 \label{fig:phase}
\end{figure}


\Fref{fig:phase} summarizes our state-of-the-art understanding on the
phase structure of QCD matter including conjectures which are not
fully established.  At present, relatively firm statements can be made
only in limited cases -- phase structure at finite $T$ with small
baryon density ($\muB\ll T$) and that at asymptotically high density
($\muB\gg\LQCD$).  Below we will take a closer look at
\fref{fig:phase} from a smaller to larger value of $\muB$ in order.


\paragraph{Hadron-quark phase transition at $\muB=0${\rm :}}

The QCD phase transition at finite temperature with zero chemical
potential has been studied extensively in the numerical simulation on
the lattice.  Results depend on the number of colours
and flavours as expected from the analysis of effective theories on
the basis of the renormalization group together with the universality
\cite{Svetitsky:1982gs,Pisarski:1983ms}.  A first-order deconfinement
transition for $\Nc=3$ and $\Nf=0$ has been established from the
finite size scaling analysis on the lattice \cite{Fukugita:1989yw},
and the critical temperature is found to be $\Tc\simeq 270\MeV$.  For
$\Nf>0$ light flavours it is appropriate to address more on the
chiral phase transition.  Recent analyses on the basis of the
staggered fermion and Wilson fermion indicate a crossover from the
hadronic phase to the quark-gluon plasma for realistic $u$, $d$ and
$s$ quark masses \cite{Aoki:2006we,DeTar:2009ef}.  The pseudo-critical
temperature $\Tpc$, which characterizes the crossover location, is
likely to be within the range $150\MeV-200\MeV$ as summarized in
\sref{sec:pseudo-CT}.

Even for the temperature above $\Tpc$ the system may be strongly
 correlated and show non-perturbative phenomena such as the existence
of hadronic modes or pre-formed hadrons in the quark-gluon plasma at
$\muB=0$ \cite{Hatsuda:1985eb,DeTar:1985kx} as well as at $\muB\neq0$
\cite{Kitazawa:2001ft,Abuki:2001be,Nishida:2005ds}.  Similar phenomena
can be seen in other strong coupling systems such as the
high-temperature superconductivity and in the BEC regime of ultracold
fermionic atoms \cite{Levin:2005}.


\paragraph{QCD critical points{\rm :}}

\begin{figure}
 \begin{center}
 \includegraphics[width=0.3\textwidth]{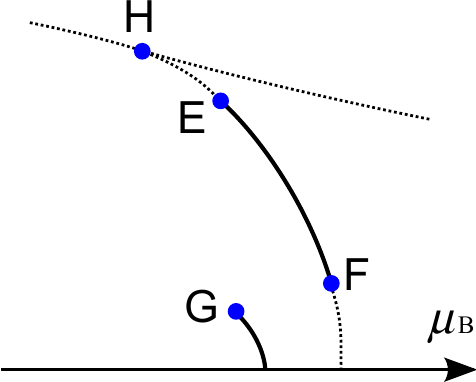}
 \end{center}
 \caption{Characteristic points on the QCD phase diagram.  \textsf{E}
   represents so-called the QCD critical point.  \textsf{F} is another
   critical point induced by the quark-hadron continuity.  \textsf{G}
   is the critical point associated with the liquid-gas transition of
   nuclear matter.  \textsf{H} refers to a region which looks like an
   approximate triple point.  See the text for details.}
 \label{fig:points}
\end{figure}

In the density region beyond $\muB\sim T$ there is no reliable
information from the first-principle lattice QCD calculation.
Investigation using effective models is a pragmatic alternative then.
Most of the chiral models suggest that there is a
\textit{QCD critical point} located at $(\muB=\muE,T=\TE)$ and the
chiral transition becomes first-order (crossover) for $\muB>\muE$
($\muB<\muE$) for realistic $u$, $d$ and $s$ quark masses
\cite{Asakawa:1989bq,Barducci:1989wi,Wilczek:1992sf,Berges:1998rc}
(see the point \textsf{E} in \fref{fig:points}).  The criticality
implies enhanced fluctuations, so that the search for the QCD critical
point is of great experimental interest
\cite{Stephanov:1998dy,Stephanov:1999zu}.

There is also a possibility that the first-order phase boundary ends
at another critical point in the lower-$T$ and higher-$\muB$ region
whose location we shall denote by $(\muF,\TF)$ as shown by the point
\textsf{F} in \fref{fig:points}.  As discussed in \sref{sec:diquark},
the cold dense QCD matter with three degenerate flavours may have no
clear border between superfluid nuclear matter and superconducting
quark matter, which is called the \textit{quark-hadron continuity}.

In reality, the fate of the above critical points (\textsf{E} and
\textsf{F}) depend strongly on the relative magnitude of the 
strange quark mass $\mst$ and the typical values of $T$ and $\muB$ at
the phase boundary.


\paragraph{Liquid-gas phase transition of nuclear matter{\rm :}}

Since the nucleon mass is $m_N\simeq 939\MeV$ and the binding energy
in isospin-symmetric nuclear matter is around $16\MeV$, a
non-vanishing baryon density of nuclear matter starts arising at
$\muB=\mu_{\rm NM}\simeq 924\MeV$ at $T=0$.  At the threshold
$\muB=\mu_{\rm NM}$, the density $\nB$ varies from zero to the normal
nuclear density $n_0=0.17\fm^{-3}$.  For $0 < \nB < n_0$ the nuclear
matter is fragmented into droplets with $\nB=n_0$, so that
$\nB<n_0$ is achieved on spatial average.  This is a typical
first-order phase transition of the liquid-gas type.  The first-order
transition weakens as $T$ grows and eventually ends up with a
second-order critical point at $(\muG, \TG)$ as indicated by the point
\textsf{G} in \fref{fig:points}.  Low energy HIC experiments indicate
that $\muG\sim\mu_{\rm NM}$ and $\TG=15\sim20\MeV$
\cite{Chomaz:2004nw}.
  

\paragraph{Quakyonic matter{\rm :}}

The Statistical Model is successful to reproduce the experimentally
observable particle abundances at various $\sqrt{s_{_{NN}}}$ and thus
various $\muB$.  This model assumes a thermally equilibrated gas of
non-interacting mesons, baryons and resonances for a given $T$ and
$\muB$ \cite{Cleymans:1998yb,Becattini:2005xt,Andronic:2008gu}.
Within the model description one can extract $T$ and $\muB$ from the
HIC data by fitting the particle ratios.  The accumulation of
extracted points makes a curve on the $\muB$--$T$ plane, which is
called the chemical freeze-out line.

The freeze-out line is not necessarily associated with any of QCD
phase boundaries.  Nevertheless, there is an argument to claim that
the sudden freeze-out of chemical compositions should take place close
to the phase transition \cite{BraunMunzinger:2003zz}.  Along the
freeze-out line the thermal degrees of freedom are dominated by mesons
for $\muB\ll m_N$.  The higher $\muB$ becomes, the more baryons are
excited.  This indicates that there must be a transitional change
at $(\TH,\muH)$, where the importance of baryons in thermodynamics
surpasses that of mesons.  This happens around
$\muH=350\sim400\MeV$ and $\TH=150\sim160\MeV$ according to the
Statistical Model analysis.

It is interesting that such a phase structure is suggested from the
large $\Nc$ limit of QCD.\ \ When $\Nc$ is large, quark loops are
suppressed by $1/\Nc$ as compared to gluon contributions
\cite{'tHooft:1973jz,Witten:1979kh}.  A finite baryon number density
arises and the pressure grows of order $\Nc$ once $\muB$ becomes
greater than the lowest baryon mass $\MB$.  Such cold dense matter
in the $\Nc=\infty$ world is named \textit{quarkyonic matter}
\cite{McLerran:2007qj}.  Then, the phase diagram of large-$\Nc$ QCD
consists of three regions separated by first-order phase transitions,
i.e.\ the confined, deconfined and quarkyonic phases.  The meeting
point of the three first-order phase boundaries is the
\textit{triple point} whose remnant for finite $\Nc$ is indicated by
the point \textsf{H} in \fref{fig:points}, as suggested in
\cite{Andronic:2009gj}.  We will revisit the idea of
quarkyonic matter in \sref{sec:quarkyonic}.


\paragraph{Colour superconductivity{\rm :}}

If $\muB$ is asymptotically large, {\rm i.e.\ } $\muB\gg\LQCD$, the
ground state of QCD matter can be analyzed in terms of the
weak-coupling methods in QCD.\ \ Also we can count on the knowledge
from condensed matter physics with quarks substituting for electrons.
In this analogue between electrons in metal and quarks in quark
matter, one may well anticipate that the ground state of QCD matter at
low $T$ should form Cooper pairs leading to colour superconductivity
(CSC)
\cite{Rajagopal:2000wf,Alford:2007xm,Huang:2010nn,Schmitt:2010pn,%
Barrois:1977xd,Bailin:1983bm}.  Theoretical characterization will be
further elucidated in \sref{sec:diquark0} and \sref{sec:diquark}.

There are many patterns of Cooper pairing and thus many different CSC
states.  The search for the most stable CSC state still remains
unsettled except for $\muq\gg\LQCD$ or $\mst\to0$.  It is in fact the
strange quark mass $\mst$ that makes the problem cumbersome.  In the
intermediate density region particularly, the Fermi surface mismatch
$\delta\muq$ of different quark flavours is given by 
$\delta\muq\sim\mst^2/\muq$.  When the gap energy $\Delta$ is
comparable with $\delta\muq$, an inhomogeneous diquark condensation
may have a chance to develop energetically.  Such an inhomogeneous CSC
state gives rise to a crystal structure with respect to $\Delta(x)$,
that is the crystalline CSC phase \cite{Alford:2000ze}.  There are, at
the same time, various candidates over the crystalline CSC phase in
this intermediate density region and the true ground state has not
been fully revealed there (see \sref{sec:inhomogeneous}).
\\

So far we have quickly looked over the key phases labelled in
\fref{fig:phase} and important points specifically picked up in
\fref{fig:points}.  In \sref{sec:confinement} we will proceed to the
theoretical framework to deal with the phase transitions of quark
deconfinement and chiral restoration, respectively.


\section{Order parameters for the QCD phase transition} 
\label{sec:confinement}

QCD has (at least) three order parameters for quark deconfinement,
chiral symmetry restoration and colour superconductivity as discussed
below.


\subsection{Polyakov loop and quark deconfinement}
\label{sec:pol-dec}

The Polyakov loop which characterizes the deconfinement transition in
Euclidean space-time is defined as
\cite{Polyakov:1978vu,Susskind:1979up}
\begin{equation}
 L(\bx) = \mathcal{P}\exp\biggl[ -\rmi g\int_0^\beta \!\rmd x_4\,
  A_4(\bx,x_4) \biggr] ,
\label{eq:Polyakov}
\end{equation}
which is an $\Nc\times\Nc$ matrix in colour space.   Here $\beta$ is
the inverse temperature $\beta=1/T$, and $\mathcal{P}$ represents the
path ordering.  We will use $\ell$ to represent the traced Polyakov
loop,
\begin{equation}
 \ell= \frac{1}{\Nc} \tr L .
\end{equation}
Let us consider the centre $\mathrm{Z}(\Nc)$ of the colour gauge group
$\mathrm{SU}(\Nc)$:  elements of the centre commute with all
$\mathrm{SU}(\Nc)$ elements and can be written as $z_k\one$ with
$z_k=\rme^{2\pi\rmi k/\Nc}$ ($k=0,1,\dots,\Nc-1$) and $\one$ being an
$\Nc\times\Nc$ unit matrix.  Under a non-periodic gauge transformation
of the following form; $V_k(x)=[z_k\one]^{x_4/\beta}$, the gauge
fields receive a constant shift,
\begin{equation}
 A_4 \;\longrightarrow\; A_4^k =
  V_k \bigl[ A_4 -(\rmi g)^{-1}\partial_4 \bigr] V_k^\dagger
  = A_4 - \frac{2\pi k}{g\Nc \beta} ,
\label{eq:transformedA}
\end{equation}
so that the traced Polyakov loop transforms as $\ell\to z_k\ell$.
Because $A_4^k$ still keeps the periodicity in $x_4$, such a
non-periodic gauge transformation still forms a symmetry of the gauge
action.  This is called centre symmetry
\cite{Svetitsky:1985ye,Svetitsky:1982gs}.  The quark action (with the
quark field denoted by $\psi$) explicitly breaks centre symmetry
because the transformed field $V_k(x)\psi(x)$ does not respect the
anti-periodic boundary condition any longer.  Thus, centre symmetry is
an exact symmetry only in the pure gluonic theory where dynamical
quarks are absent or quark masses are infinitely heavy
($\mq\to\infty$).

The expectation value of the Polyakov loop and its correlation in the
pure gluonic theory can be written as
\cite{McLerran:1981pb,Nadkarni:1986cz,Nadkarni:1986as}
\begin{eqnarray}
\label{eq:phi-L}
 && \Phi = \langle \ell(\bx) \rangle = \rme^{-\beta \fq} , \qquad
 \bar{\Phi} = \langle \ell^\dagger(\bx)\rangle
  = \rme^{-\beta \fqb} , \\
\label{eq:phi-LL}
 && \langle \ell^\dagger(\bx) \, \ell(\by) \rangle =
  \rme^{-\beta \fqq(\bxs-\bys)} .
\end{eqnarray}
Here, the constant $\fq$ ($\fqb$) independent of $\bx$ is an excess
free energy for a static quark (anti-quark) in a hot gluon medium.
\footnote{Strictly speaking, the expectation value in the pure gluonic
  theory $\langle \ell(\bx) \rangle $ should be defined by taking the
  limit $\mq\to +\infty$ after taking the thermodynamic limit
  $V \rightarrow \infty$, so that it takes a real value.
  Alternatively, one may use  $|\langle \ell(\bx) \rangle|$ to define
  the free energy of a single quark.}
Also, $\fqq(\bx-\by)$ is an excess free energy for an anti-quark at
$\bx$ and a quark at $\by$.
\footnote{Even when there are dynamical quarks, one may use the
  Polyakov loops, \eref{eq:phi-L} and \eref{eq:phi-LL}, to define the
  heavy-quark free energies.}

In the confining phase of the pure gluonic theory, the free energy of
a single quark diverges ($\fq\to\infty$) and the potential between a
quark and an anti-quark increases linearly at long distance
($\fqq(r\to\infty) \rightarrow \sigma r$ with $r=|\bx-\by|$), 
which leads to $\Phi\to 0$ and
$\langle\ell^\dagger(r\to\infty)\,\ell(0)\rangle \to 0$.  On the other
hand, in the deconfined phase, the free energy of a single quark is
finite ($\fq < \infty$).  Also the potential between a quark and an
anti-quark is of the Yukawa type at long distance with a magnetic
screening mass $m_{\rm M}$
\cite{Arnold:1995bh,Hart:2000ha,Maezawa:2010vj},
\begin{equation}
 \fqq(r\to\infty) \rightarrow 
  \fqb + \fq + \alpha\,\frac{\rme^{-m_{\rm M} r}}{r} ,
\end{equation}
where $\alpha$ is a dimensionless constant.  Note that the
  glueball exchange with mass of $O(g^2 T)$ which is smaller than
  the electric screening scale $O(gT)$ dominates over the long-range
  correlation at weak coupling.
Therefore, $\Phi\neq0$ and
$\langle\ell^\dagger(r\to\infty)\,\ell(0)\rangle \neq 0$.


\begin{table}
 \begin{tabular}{lll}
  \hline
    & Confined (Disordered) Phase & Deconfined (Ordered) Phase\\
  \hline
  Free Energy & $\fq=\infty$ & $\fq<\infty$ \\
    & $\fqq\sim\sigma r$
    & $\displaystyle \fqq\sim \fq + \fqb
      + \alpha \frac{\rme^{-m_{\rm M}r}}{r}$
  \vspace{.5em}\\
  Polyakov Loop & $\langle\ell\rangle=0$
    & $\langle\ell\rangle\neq0$ \\
  $\quad (r \to\infty)$ 
  & $\langle\ell^\dagger(r)\ell(0)\rangle\to 0$
  & $\langle\ell^\dagger(r)\ell(0)\rangle \to
    |\langle \ell \rangle |^2 \neq 0  $ \\
  \hline
 \end{tabular}
 \caption{Behaviour of the expectation value and the correlation of
   the Polyakov loop in the confined and deconfined phases in the pure
   gluonic theory.}
 \label{tab:classify}
\end{table}


The qualitative behaviour of $\Phi=\langle\ell\rangle$ and
$\langle\ell^\dagger \ell\rangle$ is summarized in
\tref{tab:classify}.  We see that $\Phi$ can nicely characterize the
state of matter in such a way that $\ell$ behaves like a magnetization
in 3D classical spin systems
\cite{Svetitsky:1982gs,Svetitsky:1985ye,Ogilvie:1983ss}.  The pure
gluonic theory for $\Nc=2$ and $3$ have been studied in lattice gauge
simulations with the finite-size scaling analysis
\cite{Okawa:1987nd,Fukugita:1989yw}.  It was shown that there is a
second-order phase transition for $\Nc=2$ and a first-order phase
transition for $\Nc=3$:  For $\Nc=2$ the critical exponent is found to
agree with the $\mathrm{Z}(2)$ Ising model in accordance with the
universality.  For $\Nc=3$ the Ginzburg-Landau free energy for the
Polyakov loop has a cubic invariant, so that the first-order
transition can natually be induced
\cite{Svetitsky:1985ye,Fukugita:1989yw,Karsch:2000xv}.  Recent studies
of the pure gluonic theory with $\Nc=4$, $6$, $8$, $10$ indicate that
the transition is of first order for $\Nc \ge 3$ and becomes stronger
as $\Nc$ increases \cite{Lucini:2003zr,Datta:2009jn}.  This behavior
is similar to that of the 3D $\Nc$-state Potts model
\cite{Wu:1982ra}.

The idea to construct the deconfinement order parameter can be
extended to a more general setup.  Suppose that there is an arbitrary
operator $\mathcal{O}$ written in terms of gauge fields that is not
invariant under centre transformation.  (If we take $\mathcal{O}$ as
the quark propagator straight up along the imaginary-time direction,
the following argument shall result in the standard definition of the
Polyakov loop.)  The expectation value of $\mathcal{O}$ in the pure
gluonic theory can be decomposed by the triality projection
\cite{Detar:1982wp};
$\langle \mathcal{O}[A]\rangle_{\rm pure} = \sum_{n=0}^{\Nc-1}
 \langle \mathcal{O}_n[A] \rangle_{\rm pure}$ with
\begin{equation}
 \langle \mathcal{O}_n[A] \rangle_{\rm pure}
  = \frac{1}{\Nc}\sum_{k=0}^{\Nc-1}
  \langle \mathcal{O}[A^k] \rangle_{\rm pure} \;
  \rme^{-2\pi\rmi kn/\Nc} ,
\end{equation}
where $A^k$ is defined in \eref{eq:transformedA}.  Then,
$\langle\mathcal{O}_n[A]\rangle_{\rm pure}$ with $n\neq0$ can be an
order parameter sensitive to the spontaneous breaking of centre
symmetry.  A particular choice $\mathcal{O}=\bar{\psi}\psi$ 
\cite{Gattringer:2006ci,Bilgici:2008qy}
 (so-called the dual condensate) has some
 practical advantages \cite{Fischer:2009wc}.

Centre symmetry discussed so far is rigorously defined only in the
pure gluonic system as already mentioned:  In the presence of
dynamical quarks, centre symmetry is broken explicitly, so that $\Phi$
and $\bar{\Phi}$ always take finite values.  This is analogous to the
spin system under external magnetic field, in which the magnetization
is always non-vanishing \cite{Gavai:1985vi}.  Nevertheless, $\Phi$ may
be estimated as $\Phi=\rme^{-\beta M}$ with a hadronic mass scale
($M\sim 0.8\GeV$) at low $T$ and thus the low-$T$ phase approximately
preserves centre symmetry.


\subsection{Chiral condensate and dynamical breaking of chiral symmetry}
\label{sec:chiral-breaking}

In the QCD vacuum at $T=\muB=0$, chiral symmetry is spontaneously
broken, which is the source of hadron masses.  It is a common wisdom
that the chiral symmetry breaking is driven by the expectation value
of an operator which transforms as $(\Nf,\Nf^\ast)+(\Nf^\ast,\Nf)$
under chiral symmetry.  A simplest choice of the order parameter for
the chiral symmetry breaking is a bilinear form called the chiral
condensate,
\begin{equation}
 \langle \bar{\psi} \psi \rangle
  = \langle \bar{\psi}_{{\rm R}} \psi_{{\rm L}} 
   + \bar{\psi}_{{\rm L}} \psi_{{\rm R}}  \rangle ,
\label{eq:chiral-condensate}
\end{equation}
where colour and flavour indices of the quark fields are to be
summed.  If the above chiral condensate is non-vanishing even after
taking the limit of zero quark masses, chiral symmetry is
spontaneously broken according to the pattern $\calG\to\calH$ with
\begin{eqnarray}
 \calG &=&
  \mathrm{SU}(\Nf)_{\rm L} \times \mathrm{SU}(\Nf)_{\rm R} \times
  \mathrm{U}(1)_{\rm B} \times \mathrm{Z}({2\Nf})_{\rm A}
\label{eq:patternG} \\
 \calH &=&
  \mathrm{SU}(\Nf)_{\rm V} \times \mathrm{U}(1)_{\rm B} ,
\label{eq:patternH}
\end{eqnarray}
which leads to $\Nf^2-1$ massless Nambu-Goldstone bosons for $\Nf>1$.
\footnote{Strictly speaking, the quotient groups
  $\calG' = \calG/[\mathrm{Z}(\Nf)\times\mathrm{Z}(\Nf)]$
  and $\calH' = \calH/\mathrm{Z}(\Nf)$ need to be introduced
  to avoid double counting of the discrete centre group operation.  A
  simplest example of this kind is
  $\mathrm{U}(N)=[\mathrm{SU}(N)\times\mathrm{U}(1)]/\mathrm{Z}(N)$
  \cite{Balachandran:2005ev}.}
The $\ua$ symmetry in the classical level of the QCD Lagrangian is
broken down explicitly to $\mathrm{Z}({2\Nf})_{\rm A}$ in the quantum
level.  Then, the $\ua$ current is no longer conserved ($\ua$
anomaly); \footnote{The anomaly relation does not necessarily
  guarantee the absence of the massless Nambu-Goldstone boson.  In the
  Schwinger model which is an exactly solvable QCD analogue, the $\ua$
  problem is resolved by the Kogut-Susskind dipole ghosts.}
\begin{equation}
 \partial_\mu j_5^\mu = -\frac{g^2\Nf}{32\pi^2}
  \epsilon^{\alpha\beta\mu\nu} F^a_{\alpha\beta}F^a_{\mu\nu} .
\end{equation}
The right-hand side of the above relation is nothing but the
topological charge density.  Thus, gauge configurations with
non-trivial topology are microscopically responsible for the $\ua$
anomaly.  In other words, the $\ua$ current could be approximately
conserved if the gauge configurations are dominated by topologically
trivial sectors.  We will come back to this point later to address
\textit{effective} restoration of $\ua$ symmetry in the medium
\cite{Shuryak:1993ee}.

Note that \eref{eq:chiral-condensate} is not a unique choice for the
order parameter
\cite{Stern:1998dy,Kogan:1998zc,Watanabe:2003xt,Harada:2009nq}. 
For example, one may consider the following four-quark condensate,
\begin{eqnarray}
 \biggl\langle \bar{\psi} \frac{\lambda^a}{2}
  (1-\gamma_5)\psi\cdot
  \bar{\psi}\frac{\lambda^a}{2}
  (1+\gamma_5)\psi \Bigr\rangle 
   =  \biggl\langle  
 \bar{\psi}_{{\rm R}} \lambda^a  {\psi}_{{\rm L}}  \cdot
 \bar{\psi}_{{\rm L}} \lambda^a  {\psi}_{{\rm R}}
  \biggr\rangle .
\label{eq:chiral-higher}
\end{eqnarray}
If this is non-vanishing, the ground state breaks chiral symmetry
$\calG$ to $\calH \times \mathrm{Z}(\Nf)_{\rm A}$ where
$\mathrm{Z}(\Nf)_{\rm A}$ corresponds to a discrete axial rotation.
If the bilinear condensate \eref{eq:chiral-condensate} is non-zero,
the four-quark condensate \eref{eq:chiral-higher} takes a finite value
in general.  However, a non-zero value of \eref{eq:chiral-higher} does
not necessarily enforce a finite value of
\eref{eq:chiral-condensate}.  In fact, if $\mathrm{Z}({\Nf})_{\rm A}$
symmetry is left unbroken, \eref{eq:chiral-condensate} must vanish.
As long as the Dirac determinant in QCD is positive definite,
possibility of unbroken $\mathrm{Z}({\Nf})_{\rm A}$ symmetry has been
ruled out by the exact QCD inequality \cite{Kogan:1998zc}.  However,
it is not necessary the case for finite $\muB$ with which the
  positivity of the Dirac determinant does not hold.
 

\subsection{Diquark condensate and colour superconductivity}
\label{sec:diquark0}

QCD at high baryon density shows a novel mechanism of spontaneous
chiral symmetry breaking \cite{Alford:1998mk}.  The fundamental
degrees of freedom in the CSC phase with three colours and three
flavours are the diquarks defined as
\begin{equation}
\hspace{-5em}
 (\varphi_{\rm L}^{\dagger})_{\alpha i}
  \sim \epsilon_{\alpha\beta\gamma}\, \epsilon_{ijk}\,
  (\psi_{\rm L}^{\rm t})_{\beta j} C (\psiL)_{\gamma k} , \quad
 (\varphi_{\rm R}^{\dagger})_{\alpha i}
  \sim \epsilon_{\alpha\beta\gamma}\, \epsilon_{ijk}\,
  (\psi_R^{\rm t})_{\beta j} C (\psiR)_{\gamma k} ,
\label{eq:LR-d}
\end{equation}
where $(i,j,k)$ are flavour indices and $(\alpha,\beta,\gamma)$ are
colour indices.  Note that the charge conjugation matrix
$C=\rmi\gamma^2\gamma^0$ is necessary to make $\varphi_{\rm L/R}$ be
Lorentz scalar.  Then, $\varphi_{\rm L/R}$ is a triplet both in colour
and flavour, so that it transforms in the same way as the quark field
$\psi_{\rm L/R}$.

Under certain gauge fixing, one may consider the expectation values of
these operators.  They are called the diquark condensates as will be
discussed further in \sref{sec:diquark}.  Instead, we can also
construct an analogue of \eref{eq:chiral-condensate} in terms of the
diquarks,
\begin{equation}
 \langle \varphi_{\rm R}^{\dagger} \varphi_{\rm L}^{\phantom{\dagger}}
  \rangle + \langle \varphi_{\rm L}^{\dagger}
  \varphi_{\rm R}^{\phantom{\dagger}} \rangle ,
\label{eq:chiral-CFL}
\end{equation}
where colour and flavour indices are summed.  This is a
gauge-invariant four-quark condensate which characterizes the
spontaneous chiral symmetry breaking in CSC \cite{Rajagopal:2000wf}.
Unlike the bilinear operator in \eref{eq:chiral-condensate}, the
four-quark operator here keeps additional invariance
$\mathrm{Z}(2)_{\rm L} \times \mathrm{Z}(2)_{\rm R}$ corresponding
to the reflections; $\psiL\to-\psiL$ and $\psiR\to-\psiR$,
 which is broken down to $\mathrm{Z}(2)$ due to the $\ua$ anomaly.

In the CSC phase, not only chiral symmetry but also
$\mathrm{U}(1)_{\rm B}$ symmetry associated with the baryon number
conservation may be spontaneously broken.   A colour singlet
order parameter to detect such symmetry breaking can be
\begin{equation}
 \epsilon_{\alpha\beta\gamma} \epsilon_{ijk}
  \bigl\langle (\varphi_{\rm L/R})_{\alpha i} \,
  (\varphi_{\rm L/R})_{\beta j} \,
  (\varphi_{\rm L/R})_{\gamma k} \bigr\rangle .
\end{equation}
The six-quark operator here breaks $\mathrm{U}(1)_{\rm B}$ with its
$\mathrm{Z}(6)_{\rm B}$ subgroup maintained.  If this condensate is
non-zero, there appears exactly massless Nambu-Goldstone boson because
the baryon number symmetry is an exact symmetry in the QCD Lagrangian.


\section{Chiral phase transition at finite temperature}
\label{sec:uncertain}

The chiral phase transition at finite $T$ with $\muB=0$ has been and
is being extensively studied by the renormalization group method near
the critical point \`{a} la Ginzburg-Landau-Wilson and by the
lattice-QCD simulations.  In this section we will briefly summarize
the current status of these studies.  (See \cite{DeTar:2009ef} for
further details.)


\subsection{Ginzburg-Landau-Wilson analysis}
\label{sec:order}

If the phase transition is of second order or of weak first order, one
may write down the free-energy functional in terms of the order
parameter field $\Phi$ as a power series of $\Phi/\Tc$.  The large
fluctuation of $\Phi$ near the critical point is then taken into
account by the renormalization group method.  This is called the
Ginzburg-Landau-Wilson approach.  For chiral phase transition in QCD,
the relevant order parameter field is a $\Nf\times\Nf$ matrix in
flavour space,
$\Phi_{ij}\sim \langle\bar{\psi}_j(1-\gamma_5)\psi_i\rangle$. 
Under the flavour chiral rotation
$\mathrm{U}(\Nf)_{\rm L}\times \mathrm{U}(\Nf)_{\rm R}$, $\Phi$
transforms as $\Phi\to V_{\rm L} \Phi V_{\rm R}^\dagger$.  Then the
Ginzburg-Landau free energy in three spatial dimensions ($D=3$) with
full $\mathrm{U}(\Nf)_{\rm L}\times \mathrm{U}(\Nf)_{\rm R}$ symmetry
up to the quartic order in $\Phi_{ij}$ becomes
\cite{Pisarski:1983ms,Wilczek:1992sf};
\begin{equation}
\hspace{-2em}
 \Omega_{\rm sym}= \frac{1}{2}\tr\nabla \Phi^\dagger \nabla \Phi
  + \frac{a_0}{2} \tr \Phi^\dagger \Phi
  + \frac{b_1}{4!}\bigl(\tr\Phi^\dagger \Phi\bigr)^2
  + \frac{b_2}{4!}\tr\bigl(\Phi^\dagger \Phi\bigr)^2 .
\label{eq:chiral-scalar}
\end{equation} 
Effects of temperature $T$ enter through the parameters $a_0$, $b_1$
and $b_2$.  Note that $\Omega_{\rm sym}$ is bounded from below as long
as $b_1+b_2/\Nf >0$ and $b_2 >0$ are satisfied.  The renormalization
group analysis of \eref{eq:chiral-scalar} on the basis of the
leading-order $\epsilon (=4-D)$ expansion leads to a conclusion that
there is no stable IR fixed point for $\Nf>\sqrt{3}$
\cite{Pisarski:1983ms}.  This implies that the thermal phase
transition described by \eref{eq:chiral-scalar} is of the
fluctuation-induced first order for two or more flavours.

In QCD, however, there is $\ua$ anomaly and the correct chiral
symmetry is
$\mathrm{SU}(\Nf)_{\rm L}\times\mathrm{SU}(\Nf)_{\rm R}
\times \mathrm{U}(1)_{\rm B} \times \mathrm{Z}({2\Nf})_{\rm A}$ for
$\Nf$ massless quarks.  The lowest dimensional operator which breaks
$\ua$ symmetry explicitly while keeping the rest of chiral symmetry is
the Kobayashi--Maskawa--'t~Hooft (KMT) term
\cite{Kobayashi:1970ji,Kobayashi:1971qz,'tHooft:1976up,'tHooft:1976fv};
\begin{equation}
 \Omega_{\rm anomaly} = - \frac{c_0}{2} \,
  \bigl( \det\Phi + \det\Phi^\dagger \bigr) .
\label{eq:det-anomaly}
\end{equation}
The coefficient $c_0$, which is $T$-dependent in general, dictates the
strength of $\ua$ anomaly.  In the instanton picture
\cite{'tHooft:1976up,'tHooft:1976fv}, $c_0(T=0)$ is proportional to
the instanton density $n_{\rm inst}$, which is perturbatively
  evaluated as \cite{Gross:1980br}
\begin{equation}
 c_0(T=0) \propto n_{\rm inst}(\rho,T=0)
  = \Bigl(\frac{8\pi^2}{g^2}\Bigr)^{2\Nc} \rme^{-8\pi^2/g^2} \rho^{-5} ,
\label{eq:inst-density}
\end{equation}
with $\rho$ being a typical instanton size.

For $\Nf=3$ the KMT term becomes a cubic invariant in the order
  parameter.  Hence,
  $\Omega[\Phi] = \Omega_{\rm sym}+ \Omega_{\rm anomaly}$ leads to the
  chiral phase transition of first order.
For $\Nf=2$, on the other hand, the KMT term becomes a quadratic
invariant.  Also the chiral symmetry in this case is
 $\mathrm{SU}(2)_{\rm L}\times\mathrm{SU}(2)_{\rm R} \cong\mathrm{SO}(4)$.
Such an effective theory with $\mathrm{O}(4)$ symmetry has a
Wilson-Fisher type IR fixed point as long as the coefficient of the
quartic term of $\Phi$ is positive.  Therefore, \textit{if} the chiral
phase transition of massless $\Nf=2$ QCD is of second order, its
critical exponents would be the same as those in the 3D
$\mathrm{O}(4)$ effective theory according to the notion of
universality.
In Table \ref{tab:chiral-order} we summarize the
Ginzburg-Landau-Wilson analysis from the chiral effective theory
\cite{Pisarski:1983ms}.


\begin{table}
 \begin{center}
 \begin{tabular}{lcc}
 \hline
  & $\Nf=2$ & $\Nf\ge3$ \\
 \hline \hline
  $\ua$ symmetric ($c_0=0$) &
   Fluctuation-induced 1st order & 1st order \\
  $\ua$ broken ($c_0\neq 0$) &
   2nd order [$\mathrm{O}(4)$ universality] & 1st order \\
 \hline
 \end{tabular}
 \end{center}
\caption{Order of the chiral phase transition conjectured from the
    chiral effective theory with massless $\Nf$ flavours with and
    without the $\ua$ anomaly.}
\label{tab:chiral-order}
\end{table}



\begin{figure}
 \begin{center}
 \includegraphics[width=0.4\textwidth]{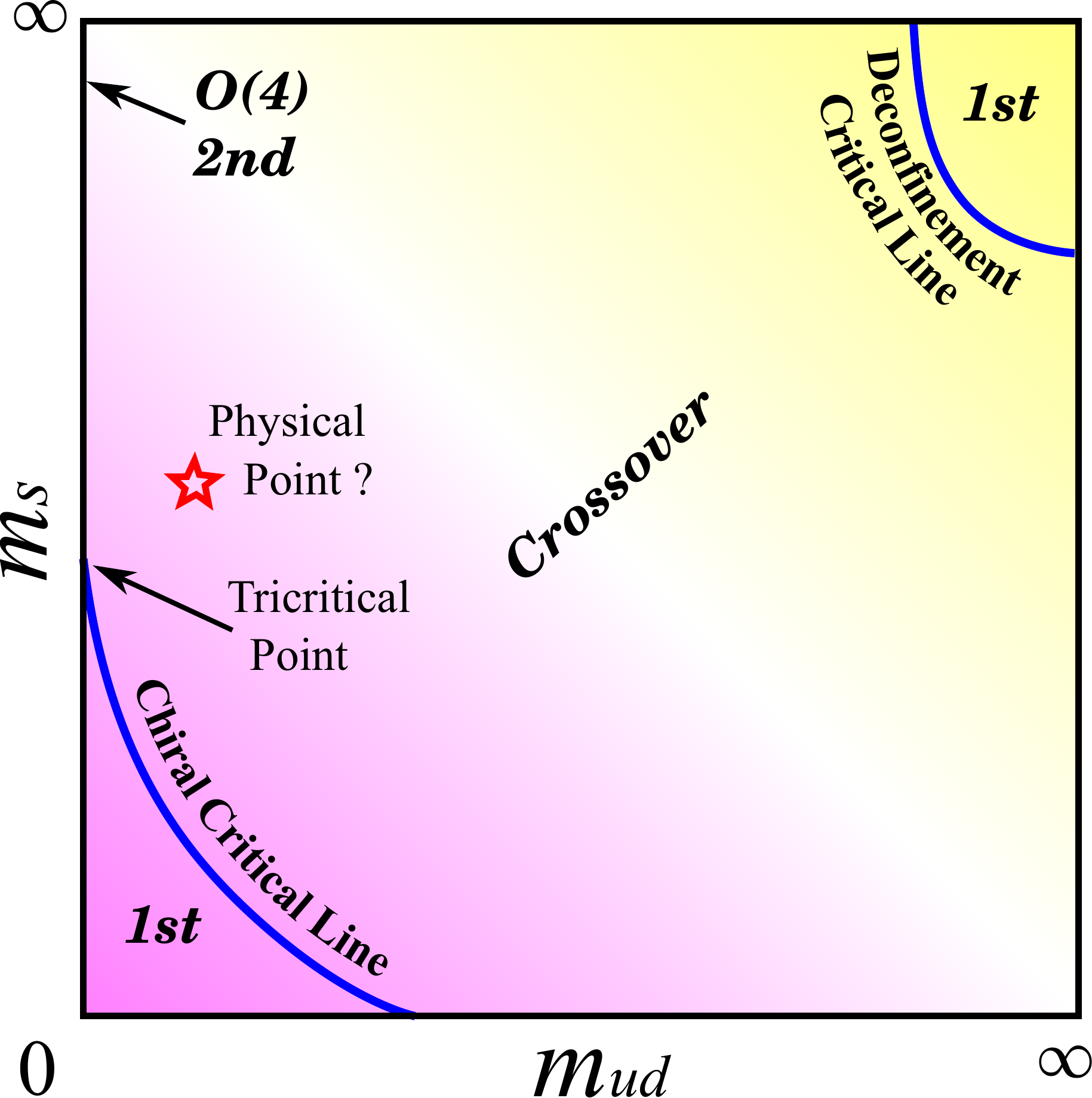}
 \end{center}
 \caption{Schematic figure of the Columbia phase diagram in
   $3$-flavour QCD at $\muB=0$ on the plane with the light and heavy
   quark masses.  The $\ua$ symmetry restoration is not taken into
   account.  Near the left-bottom corner the chiral phase transition
   is of first order and turns to smooth crossover as $\mud$ and/or
   $\mst$ increase.  The right-top corner indicates the deconfinement
   phase transition in the pure gluonic dynamics.}
 \label{fig:columbia}
\end{figure}


In the real world, none of quark is exactly massless:  For example,
$\mup = (1.5$--$3.3)\MeV$, $\mdw = (3.5$--$6.0)\MeV$ and
$\mst=105^{+ 25}_{-35}\MeV$ at the renormalization scale of $2\GeV$
\cite{Amsler:2008zzb}.  Therefore, it is useful to draw a phase
diagram by treating quark masses as external parameters.  This is
called the Columbia plot \cite{Brown:1990ev} as shown in
\fref{fig:columbia} where the isospin degeneracy is assumed
($\mup = \mdw \equiv \mud$).  The first-order chiral transition and
the first-order deconfinement transition at finite $T$ are indicated
by the left-bottom region and the right-top region, respectively.  The
chiral and deconfinement critical lines, which separate the
first-order and crossover regions, belong to a universality class of
the 3D $\mathrm{Z}(2)$ Ising model except for special points at
$\mud=0$ or $\mst=0$ \cite{Gavin:1993yk}.

If the chiral transition is of second order for massless $\Nf=2$ case,
the $\mathrm{Z}(2)$ chiral critical line meets the $\mud=0$ axis at
$\mst = \mst^{\rm tri}$ (\textit{tricritical point}) and changes its
universality to $\mathrm{O}(4)$ for $\mst > \mst^{\rm tri}$
\cite{Hatta:2002sj}.  The tricritical point at $\mst=\mst^{\rm tri}$
is a Gaussian fixed point of the 3D $\phi^6$ model (that is, the
  critical dimension is not $4$ but $3$ at the tricritical point),
so that the critical exponents take the classical (mean-field) values
\cite{Riedel-Wegner}, which is confirmed in numerical studies of the
chiral model \cite{Schaefer:2006ds}.


\begin{figure}
 \begin{center}
 \includegraphics[width=0.6\textwidth]{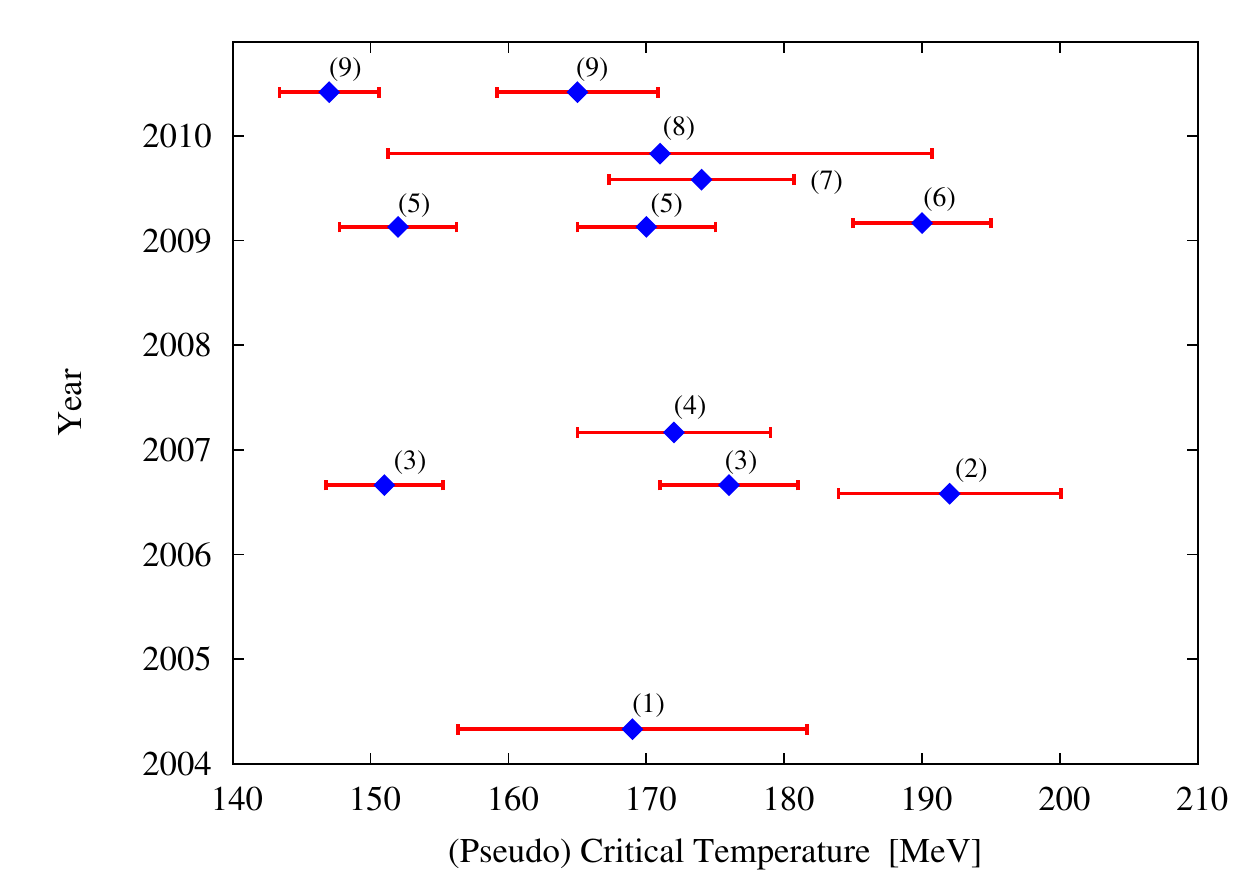}
 \end{center}
 \caption{Determination of the pseudo-critical temperature $\Tpc$ for
   thermal QCD transition(s) from recent lattice QCD simulations.
   (1) $169(12)(4)\MeV$ for $2+1$ flavours in the asqtad action with
   $N_t$ up to $8$ determined by $\chi_m/T^2$ (where $\chi_m$ is the
   chiral susceptibility) \cite{Bernard:2004je}.
   (2) $192(7)(4)\MeV$ for $2+1$ flavours in the p4fat3 staggered
   action with $N_t$ up to $6$ determined by $\chi_m$ and $\chi_L$
   (where $\chi_L$ is the Polyakov loop susceptibility)
   \cite{Cheng:2006qk}.
   (3) $151(3)(3)\MeV$ and $176(3)(4)\MeV$ for $2+1$ flavours in the
   stout-link improved staggered action with $N_t$ up to $10$
   determined by $\chi_m/T^4$ and $\chi_L$ respectively
   \cite{Aoki:2006br}.
   (4) $172(7)\MeV$ for $2$ flavours in clover improved Wilson action
   with $N_t$ up to $6$ determined by $\chi_L$ \cite{Maezawa:2007fd}.
   (5) $152(3)(3)\MeV$ and $170(4)(3)\MeV$ for $2+1$ flavours in the
   stout-link improved staggered action with $N_t$ up to $12$
   determined by $\chi_m/T^2$ and $\chi_L$ respectively
   \cite{Aoki:2009sc}.
   (6) $185$--$195\MeV$ for $2+1$ flavours in the asqtad and p4
   actions with $N_t$ up to $8$ determined by $\chi_m$ and $\chi_L$
   \cite{Bazavov:2009zn}.
   (7) $174(3)(6)\MeV$ for $2$ flavours in the improved Wilson action
   with $N_t$ up to $12$ determined by $\chi_m$ and $\chi_L$
   \cite{Bornyakov:2009qh}.
   (8) $171(10)(17)\MeV$ for $2+1$ flavours in the domain-wall action
   with $N_t=8$ determined by $\chi_m/T^2$ \cite{Cheng:2009be}.
   (9) $147(2)(3)\MeV$ and $165(5)(3)\MeV$ for $2+1$ flavours in the
   stout-link improved staggered action with $N_t$ up to $16$
   determined by $\chi_m/T^4$ and $\chi_s/T^2$ (where $\chi_s$ is the
   strange-quark susceptibility) \cite{Borsanyi:2010bp}.}
 \label{fig:lat-tc}
\end{figure}



\subsection{Lattice QCD simulations}
\label{sec:pseudo-CT}

Although the critical properties expected from the
Ginzburg-Landau-Wilson analysis discussed above are expected to be
universal, the quantities such as the critical temperature and the
equation of state depend on the details of microscopic dynamics.  In
QCD, only a reliable method known for microscopic calculation is the
lattice-QCD simulation in which the functional integration is carried
out on the space-time lattice with a lattice spacing $a$ and the
lattice volume $V$ by the method of importance sampling.  In
lattice-QCD simulations there are at least two extrapolations required
to obtain physical results; the extrapolation to the continuum limit
($a \to 0$) and the extrapolation to the thermodynamic limit
($V \to \infty$).  Therefore, lattice results receive not only
statistical errors due to the importance sampling but systematic
errors due to the extrapolations also.

For nearly massless fermions in QCD, there is an extra complication to
reconcile chiral symmetry and lattice discretization;  the Wilson
fermion and the staggered fermion have been the standard ways to
define light quarks on the lattice, while the domain-wall fermion and
the overlap fermion recently proposed have more solid theoretical
ground although the simulation costs are higher.  For various
applications of lattice-QCD simulations to the system at finite $T$ and
$\muB$, see a recent review \cite{DeTar:2009ef}.

Here we mention only two points relevant to the discussions below:
(i) The thermal transition for physical quark masses is likely to be
crossover as indicated by a star-symbol in \fref{fig:columbia}.  This
is based on the finite-size scaling analysis using staggered fermion
\cite{Aoki:2006we}.  Confirmation of this result by other fermion
formalisms is necessary, however.
(ii) The (pseudo)-critical temperature $\Tpc$ with different types of
fermions and with different lattice spacings are summarized in
\fref{fig:lat-tc}.  In view of these data with error bars, we adopt a
conservative estimate at present; $\Tpc=150$--$200\MeV$.
\footnote{Possible uncertainties in $\Tpc$ stem from discretization
  errors, conversion from the lattice unit to the physical unit, and
  the prescription of defining $\Tpc$.  For crossover transition
  $\Tpc$ may depend on which  susceptibility (either chiral
  susceptibility $\chi_m(T)$ or Polyakov loop susceptibility
  $\chi_L(T)$) are used, and also depend on $T$-dependent
  normalization of the susceptibilities.}
It has been clarified recently that improvement of the staggered
action with less taste-symmetry breaking favours smaller value of
$\Tpc\lesssim 170\MeV$ \cite{Borsanyi:2010bp,Bazavov:2010sb}.


\section{Chiral phase transition at finite baryon density}
\label{sec:chiral}

\begin{figure}
 \begin{center}
 \includegraphics[width=0.45\textwidth]{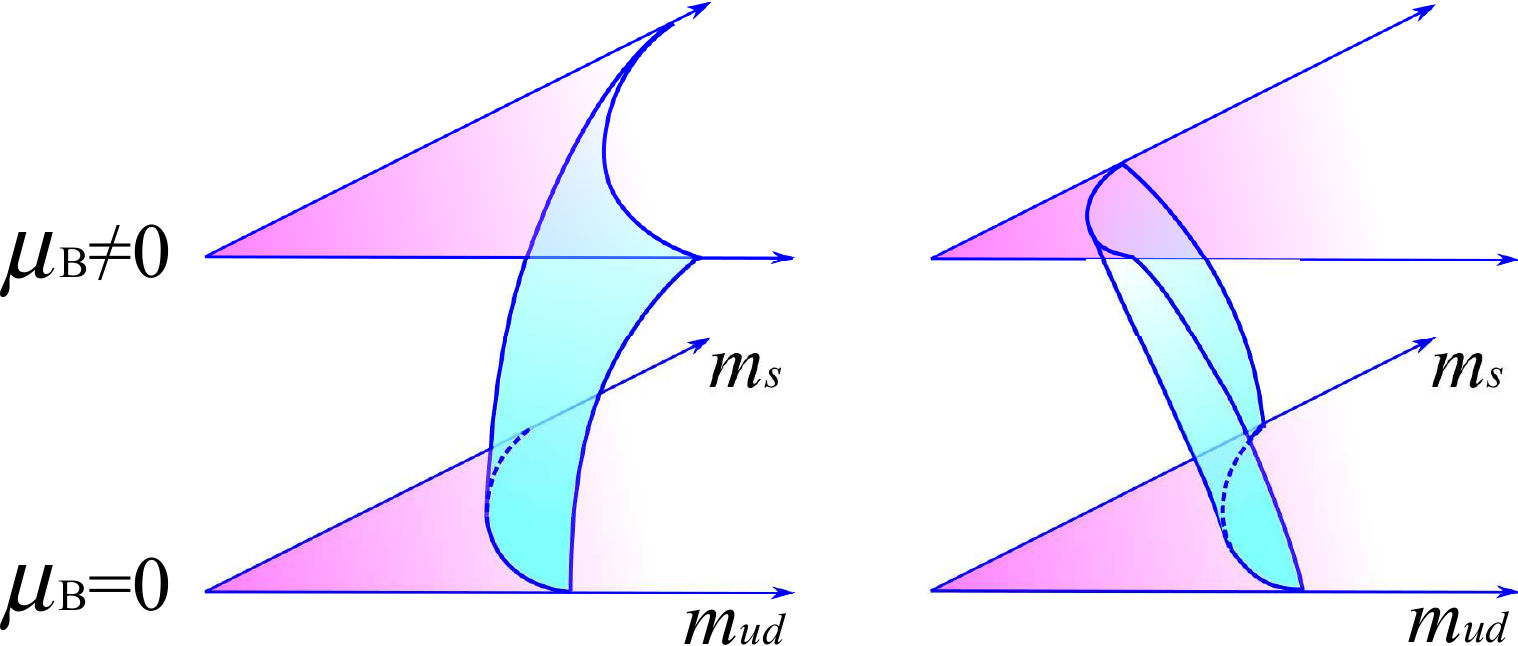}
 \end{center}
 \caption{Schematic evolution of the Columbia plot with increasing
   $\muB$ in the standard scenario (left) and the exotic scenario
   (right).}
 \label{fig:curve}
\end{figure}

Let us now introduce the baryon chemical potential $\muB$ as an extra
axis to the Columbia plot.  In the so-called ``standard scenario'' the
first-order region in the lower left corner of the Columbia plot is
elongated with increasing $\muB$ as written in the left panel of
\fref{fig:curve}.  If the physical point at $\muB=0$ is in the
crossover region, there arises a critical chemical potential $\muE$ so
that the system shows a first-order transition for $\muB > \muE$.
That is, we find a QCD critical point at $(\muE,\TE)$ on the QCD phase
diagram in the $\muB$--$T$ plane.  In the right of \fref{fig:curve},
the so-called ``exotic scenario'' is sketched, in which the size of
the first-order region shrinks as $\muB$ increases.  In this case, if
the physical point at $\muB=0$ is in the crossover region, it stays
crossover for finite $\muB$, so that there arises no critical point
(at least for small $\muB$) in the QCD phase diagram in the
$\muB$--$T$ plane.  In general the critical surface in
\fref{fig:curve} can have more complicated structure, which may allow
for several QCD critical points in the $\muB$--$T$ plane.


\subsection{Lattice QCD at low baryon density}

In the presence of $\muq\neq0$, the QCD partition function on the
lattice is written as
\begin{equation}
 Z(T,\mu) = \int [\rmd U]  \det[F(\muq)] \exp^{-\beta S_{\rm YM}(U)} ,
\label{eq:partition}
\end{equation}
where $U$ is the matrix-valued gauge field defined on the links.  The
Yang-Mills action is denoted by $\beta S_{\rm YM}(U)$ with
$\beta=2\Nc/g^2$, while the quark contribution is denoted by the
determinant of $F(\muq)=D(\muq)+\mq$ with $D(\muq)$ being the
Euclidean Dirac operator.  The quark mass $\mq$ is real and positive.
Let $\psi$ be an eigenfunction of $D(\muq)$ as
$D(\muq)\psi_i = \lambda_i\psi_i$ with an eigenvalue $\lambda_i$.  If
$\muq=0$ then $D(0)$ is an anti-Hermitean operator and thus
$\lambda_i$ should be pure imaginary.  Besides, $\gamma^5\psi_i$ is an
independent eigenfunction with an eigenvalue
$-\lambda_i=\lambda_i^\ast$, because of $\gamma^5$-Hermiticity
$\gamma^5 D(0) \gamma^5 = -D(0) = D^{\dagger}(0)$.  Thus, we always
have complex-conjugate pairs $(\lambda_i,\lambda_i^\ast)$ and the
determinant becomes real and positive;
\begin{equation}
 \det[D(0)+\mq] = \prod_i(\lambda_i+\mq)(\lambda_i^\ast+\mq) > 0 .
\label{eq:positivity}
\end{equation}
In this way, at $\muq=0$, the integrand in \eref{eq:partition} is
positive definite and the importance sampling method works fine to
evaluate the functional integral.

For $\muq\neq 0$ the $\gamma^5$-Hermiticity is replaced by
\begin{equation}
 \gamma^5 D(\muq) \gamma^5 = D^\dagger(-\muq^\ast) .
\label{eq:g5hermite}
\end{equation}
Therefore $\det [D(\muq)+\mq]$ is no longer real unless $\muq$ is pure
imaginary.  This requires us to deal with a severe cancellation of
positive and negative numbers to evaluate the functional integral.
This is the notorious \textit{sign problem} whose difficulty grows
exponentially as the lattice volume $V$ increases.

In the following we shall briefly look over some of approaches in the
lattice-QCD simulations at finite $\muq$.  For more details, see the
reviews \cite{Muroya:2003qs,Ejiri:2008nv,Lombardo:2009tf} and
references therein.

\begin{itemize}

\item
\textit{Multi-parameter reweighting method} ---
An expectation value of an operator $\mathcal{O}$ at finite $\muq$ can
be formally rewritten in terms of an ensemble average at $\muq=0$;
\begin{equation}
 \langle\mathcal{O}\rangle_{\muq}
  = \langle\mathcal{O}\cdot R(\muq)
  \rangle_{\muq=0} \langle R(\muq) \rangle_{\muq=0}^{-1} ,
\label{eq:Glasgow}
\end{equation}
where $R(\muq)\equiv {\det F(\muq)}/{\det F(0)}$ is called the
reweighting factor \cite{Barbour:1998jx}.  The gauge configurations
generated at $\muq=0$  only occasionally sample the region where
$\mathcal{O}\cdot R(\muq)$ and $R(\muq)$ are large (overlap
  problem).  Also, they take complex values (sign problem).
Therefore, such a simulation works only when   $\muq$ and $V$ are
small.

 
\begin{figure}
 \begin{center}
 \includegraphics[width=0.6\textwidth]{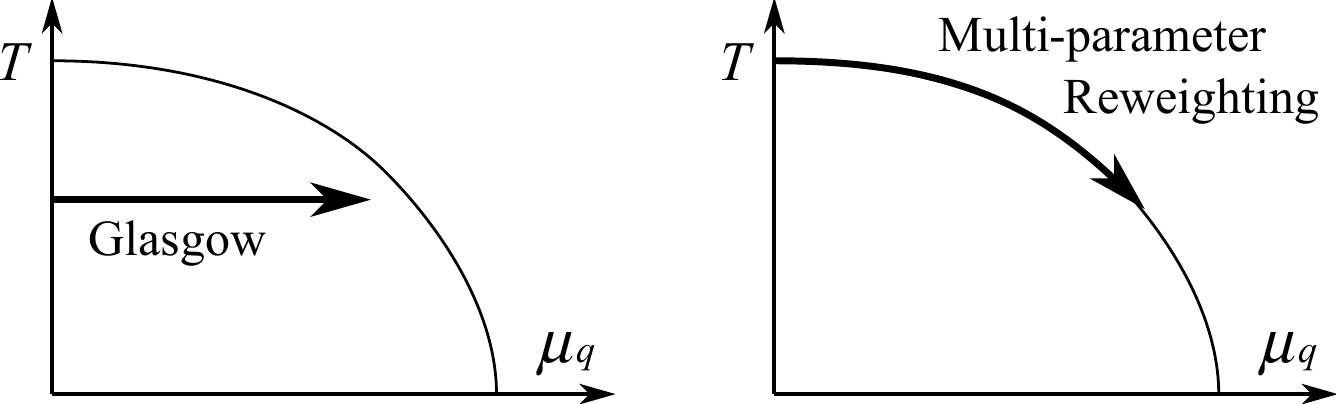}
 \end{center}
 \caption{Conceptual illustration of the single-parameter reweighting
   method (Glasgow method) in the $\muq$ direction  only and the
   multi-parameter reweighting method in  the both $\muq$ and
     $\beta$ (i.e.\ $T$) directions}.
 \label{fig:glasgow}
\end{figure}


To have a better overlap, the reweighting may be generalized towards
not only  the $\muq$ direction but also  the $\beta$ direction
 \cite{Fodor:2001pe}:
\begin{equation}
 \langle\mathcal{O}\rangle_{\muq}
 = \langle \mathcal{O}\cdot R(\muq,\delta\beta)
  \rangle_{\beta_0\!,\, \muq=0}
  \langle R(\muq,\delta\beta)
  \rangle_{\beta_0\!,\, \muq=0}^{-1} ,
\end{equation}
where the multi-parameter reweighting factor is defined as
$R(\muq,\delta\beta)\equiv R(\muq) \exp(-\delta\beta S)$ with
$\delta\beta = \beta-\beta_0$.  Here $\beta_0$ should be chosen to
maximize the overlap.  The physical temperature $T$ and
$\beta=2\Nc/g^2$ are implicitly related through the lattice spacing
$a$.  Then, the reweighting along the phase boundary in  the
$\muq$--$T(\beta)$ plane (as sketched in the right panel of
\fref{fig:glasgow}) would have a better chance to probe into larger
$\muq$.
\vspace{1em}

\item
\textit{Taylor expansion method} ---
The full evaluation of the reweighting factor $R(\muq)$ is not an easy
task even on the computer.  If $\muq$ is small enough, the right-hand
side of \eref{eq:Glasgow} can be expanded in terms of $\muq/T$
\cite{Allton:2002zi,Allton:2005gk,Gavai:2008zr},
\begin{eqnarray}
 & \langle\mathcal{O}\rangle_{\muq} = \sum_{n=0}^\infty
  c_n \Bigl(\frac{\muq}{T}\Bigr)^n , \nonumber\\
 & c_n = \frac{T^n}{n!} \frac{\partial^n}{\partial \muq^n}
  \bigl( \langle\mathcal{O}\cdot R(\muq) \rangle_{\muq=0}
  \langle R(\muq) \rangle_{\muq=0}^{-1} \bigr) .
\end{eqnarray}
The coefficients $c_n$ are written in terms of the quark propagator
and can be simulated at $\muq=0$.  Because the expansion is based on
\eref{eq:Glasgow}, the original overlap problem translates into the
convergence problem in higher order terms;  the Taylor expansion makes
sense for $\muq/T$ within a radius of convergence dictated by the
singularity closest to the origin in the complex $\muq/T$-plane.
\vspace{1em}

\item
\textit{Imaginary chemical potential method} ---
If $\muq$ is pure imaginary (which is here denoted by $\muqim$),
\eref{eq:g5hermite} reduces to the $\gamma^5$-Hermiticity, so that
there is no sign problem \cite{Alford:1998sd,deForcrand:2002ci}.
Then, it is possible to perform lattice simulations to find physical
observables by  the analytic continuation back to the real chemical
potential,
\begin{equation}
 \langle\mathcal{O}\rangle_{\muqim} = \sum_{n=0}^\infty c_n
  \Bigl(\frac{\muqim}{\mu_0}\Bigr)^n
 \:\rightarrow\:
 \langle\mathcal{O}\rangle_{\muq} = \sum_{n=0}^\infty c_n
  \Bigl(\frac{-\rmi\muq}{\mu_0}\Bigr)^n .
\end{equation}
The applicability of this method is bounded by singularities or
periodicity.  For imaginary chemical potential, $\muqim/T$ plays a
role of an angle variable which has na\"{i}ve periodicity  by $2\pi$.
However, the partition function is a function of $\muqim+gA_4$ and a
change in $\muqim$ by $2\pi/\Nc$ can be absorbed by the centre
transformation \eref{eq:transformedA}.  Therefore, the actual period
of $\muqim/T$  is $2\pi/\Nc$ (Roberge-Weiss (RW) periodicity)
\cite{Roberge:1986mm}.  In the deconfinement phase at high $T$,
especially, there is a first-order phase transition at a half of the
RW point; $\muqim/T=\pi/\Nc\sim 1$ for $\Nc=3$.  Thus the method of
imaginary chemical potential works up to $\muqim/T\lesssim 1$ at best,
which is also confirmed in model studies \cite{Sakai:2008um}.
\vspace{1em}

\item
\textit{Canonical ensemble method} ---
In the thermodynamic limit the canonical ensemble with fixed
particle number $\Nq$ is equivalent with the grand canonical ensemble
with  fixed chemical potential $\muq$ (and the mean value of
$\Nq$ is specified).  To convert  the grand canonical to the
canonical description, the Fourier transform in terms of the imaginary
chemical potential is necessary
\cite{Miller:1986cs,Hasenfratz:1991ax,Engels:1999tz,%
Alford:1998sd,Kratochvila:2005mk,Li:2008fm},
\begin{equation}
 \langle\mathcal{O}\rangle_{\Nq} = \int_0^{2\pi} \frac{\rmd\phi}{2\pi}\;
  \rme^{-\rmi \Nq\phi} \; \langle\mathcal{O}\rangle_{\muqim=\phi T} .
\end{equation}
In this case, the difficulty of the sign problem is transferred to the
integration with respect to $\phi=\muqim/T$.  This canonical approach
works fine as long as the volume is not large, for which centre
symmetry is forced to be restored for $\Nq$ that is a multiple of
$\Nc$ \cite{Detar:1982wp}.  It is a highly delicate procedure to take
the correct thermodynamic limit $V\to\infty$ in which not $\Nq$ but
$\nq=\Nq/V$ should be kept fixed.  Especially, the order of the phase
transition is sensitive to how the thermodynamic limit is approached
\cite{Fukushima:2002bk}.
\vspace{1em}

\item
\textit{Density of states method} ---
There are several variants of the density of states method depending
on the choice of a variable to rewrite the partition function.  Here
we take an example of a phase $\theta$ of the Dirac determinant
\cite{Gocksch:1988iz,Ejiri:2007ga}.  (The plaquette $P$ or energy $E$
is useful as well
\cite{Anagnostopoulos:2001yb,Ambjorn:2002pz,Fodor:2007vv}.)  The
density of states in this case reads
\begin{equation}
 \rho(\theta) = \langle \delta(\theta-\theta(U)) \rangle_{\rm pq} ,
\label{eq:dos}
\end{equation}
where the expectation value is taken by the phase quenched simulation
in which $\det F(\muq)$ is replaced by $| \det F(\muq)|$ to avoid the
sign problem.  Using $\rho(\theta)$ one can express the expectation
value of an operator $\mathcal{O}$ as
\begin{equation}
 \langle\mathcal{O}\rangle_{\muq} = \frac{1}{Z}\int \rmd\theta \,
  \rho(\theta)\, \rme^{\rmi\theta} \langle\mathcal{O}\rangle_{\theta} ,
\end{equation}
with $Z = \int\rmd\theta\, \rho(\theta)\, \rme^{\rmi\theta}$ and
\begin{equation}
  \langle\mathcal{O}\rangle_{\theta} = \frac{1}{\rho(\theta)}
  \langle \mathcal{O}\cdot \delta(\theta-\theta(U))\rangle_{\rm pq} .
\end{equation}
Both $\rho(\theta)$ and $\langle\mathcal{O}\rangle_\theta$ are
quantities calculable  in the lattice simulation.  The difficulty
of the sign problem is now translated into the precise determination
of the density of states;  \eref{eq:g5hermite} implies that the phase
quenched simulation at $\muq\neq0$ with two degenerate flavours is
identical to the simulation at finite isospin chemical potential
$\muI(=\muq)$;
\begin{equation}
 |\det F(\muq)|^2 = \det F(\muq) \det F(-\muq) .
\end{equation}
Then, the phase quenched expectation value goes through a qualitative
change as $\muq$ increases;  an exotic phase with pion condensation
appears for $\muq > m_\pi/2$ \cite{Son:2000xc}.  For small $\muq$,
both lattice-QCD simulation \cite{Ejiri:2007ga} and an analysis in the
$\chi$PT \cite{Splittorff:2007zh,D'Elia:2009tm,Lombardo:2009aw} show
that $\rho(\theta)$ behaves as Gaussian, while $\rho(\theta)$ becomes
Lorenzian for $\muq>m_\pi/2$ in the $\chi$PT.\ \ In the latter case,
extremely precise determination of $\rho(\theta)$ and
$\langle\mathcal{O}\rangle_\theta$ is crucially important.  So far,
the density of states method is applied for relatively small values of
$\muq/T$ combined with the Taylor expansion of the fermion determinant
\cite{Ejiri:2007ga,Ejiri:2009hq}.
\vspace{1em}

\item
\textit{Complex Langevin method} ---
The sign problem arises from the importance sampling to deal with the
multi-dimensional functional integral.  Therefore, it may not appear
in different quantization schemes other than the functional integral.
The stochastic quantization
\cite{Parisi:1980ys,Parisi:1984cs,Klauder:1983sp,Damgaard:1987rr,%
Namiki:1992wf} formulated in terms of the Langevin dynamics with a
fictitious time is a promising candidate for such an alternative.  If
we consider a scalar field theory defined by an action $S[\phi]$, the
Langevin equation reads
\begin{equation}
 \frac{\partial \phi(x,s)}{\partial s}
  = -\frac{\delta S[\phi]}{\delta\phi(x,s)} + \eta(x,s) ,
\end{equation}
where $s$ is a fictitious time and $\eta$ a Gaussian noise;
$\langle\eta(x,s)\rangle=0$ and
$\langle\eta(x,s)\eta(x',s')\rangle = 2\delta(x-x')\delta(s-s')$.  The
expectation value is taken over the $\eta(x,s)$ distribution and the
equilibrated value is obtained in the $s\to\infty$ limit.  At finite
$\muq$ the action and noise are complex.  The complex Langevin
dynamics can evade the sign problem and correctly describe the system
at finite density for some simple models, while there are also cases
where it may fall into a wrong answer
\cite{Aarts:2009uq,Aarts:2010gr}.  Whether this method is applicable
to dense QCD or not is still an open question.

\end{itemize}

It would be an important milestone if the lattice-QCD simulation can
show the existence/non-existence of the QCD critical point in
$\muB$--$T$ plane.  In \cite{Fodor:2001pe,Fodor:2004nz} the location
of the Lee-Yang zero in the complex lattice-coupling
($\beta=2\Nc/g^2$) plane has been investigated using the
multi-parameter reweighting for $2+1$-flavour staggered fermions.  If
the thermal transition is of first order, the Lee-Yang zero nearest to
the real axis in a finite lattice box is expected to approach to the
real axis as the lattice volume increases, while there is no such
tendency for the crossover transition.  It was then concluded that the
QCD critical point is located at
$(\muB,T)=(\muE,\TE)=(360\pm40\MeV,162\pm2\MeV)$ for physical quark
masses \cite{Fodor:2004nz}.  It has been argued, however, in
\cite{Ejiri:2005ts} that the sign problem at finite baryon density can
fool the Lee-Yang zero near the real axis.  It is also suggested to
study the behaviour of a set of Lee-Yang zeros in the complex $\beta$
plane to make a firm conclusion.

If we \textit{assume} that the convergence of the Taylor expansion in
terms of $\muq/T$ is dictated by the singularity at the critical
point, the radius of convergence is an indicator of the location of
the critical point.  In \cite{Gavai:2008zr} it was reported that the
coefficients up to sixth order yield $\TE/\Tpc=0.94\pm0.01$ and
$\muE/\Tpc=1.8\pm0.1$ for $2$-flavour staggered fermions with
$m_{\pi}/m_{\rho}=0.3$.  In the same way, with $2+1$-flavour staggered
fermions, the radius of convergence is discussed in
\cite{Miao:2008sz}, though the results are not conclusive.  The idea
of the radius of convergence has been also tested in a chiral
effective model coupled with the Polyakov loop, in which the exact
location of the critical point and the higher-order coefficients are
calculable in the mean-field approximation \cite{Wambach:2009ee}.  The
result  from the model test suggests that the relation between the
convergence radius of the expansion and the location of the critical
point is not clear.
 
In \cite{Ejiri:2008xt}  the canonical partition function is
extracted using the density of states method with an assumption of the
Gaussian distribution of the phase $\theta$ of the Dirac determinant.
In this case the standard S-shape curve in the $\muB$--$\nB$ plane is
a signature of the first-order transition.  The location of the
critical point is estimated to be $\TE/\Tpc\approx 0.76$ and
$\muE/\Tpc\approx 7.5$ for $2$-flavour staggered fermions with
$m_{\pi}\simeq 770\MeV$.  In \cite{Li:2008fm} the canonical ensemble
method is used to study the S-shape and it was found
$\TE/\Tpc=0.94\pm0.03$ and $\muE/\Tpc=3.01\pm0.12$ for $3$-flavour
clover fermions with $m_{\pi}\simeq 700\MeV$.

As shown in \fref{fig:curve} one may consider the critical surface
$\muB = \muB(\mud, \mst)$ and its behaviour near $\muB =0$ to check
whether the critical point is located in small $\muB$ region.  In
particular, on the $\mathrm{SU}(3)_{\rm V}$ symmetric line
($\mud=\mst$), one can define the critical mass as a function of
$\muq$ and make a Taylor expansion,
\begin{equation}
 \frac{\mc(\muq)}{\mc(0)} = 1
  + c_2\Bigl(\frac{\muq}{\pi \Tpc}\Bigr)^2
  + c_4\Bigl(\frac{\muq}{\pi \Tpc}\Bigr)^4 + \cdots .
\label{eq:c24}
\end{equation}
The sign of $c_2$ and $c_4$ determines whether the critical surface
behaves like the standard scenario or the exotic scenario in
\fref{fig:curve}.  Simulations with $3$-flavours of staggered fermions
on a coarse lattice ($N_t=4$) with the imaginary chemical potential
method show $c_2=-3.3(3)$ and $c_4=-47(20)$ \cite{deForcrand:2008vr},
which is consistent with the exotic scenario in \fref{fig:curve}.
This does not, however, exclude a possibility that the critical point
exists for large value of $\muq/T$.  Confirmation of this result with
different fermion formulations with smaller lattice spacing is
necessary to draw a firm conclusion \cite{deForcrand:2008zi}.

In short summary,  in any method,  the validity of QCD
simulations at finite baryon density is still limited in the region
$\muq/T<1$ at present because of the sign problem.  Although some of
the lattice-QCD simulations suggest the existence of the QCD critical
point in $\muB$--$T$ plane, the results are to be taken with care
if it is predicted at large $\muq/T$.


\subsection{Effective $\ua$ symmetry restoration}

For $\Nf=3$ the first-order transition at finite $T$ is driven by the
trilinear chiral condensates from the $\ua$-breaking KMT-type
interaction with the strength $c_0$ as discussed in
\sref{sec:order}.  In the instanton calculation $c_0$ is
proportional to the instanton density \eref{eq:inst-density} which is
screened by the medium at high $T$ and $\muq$ as
\begin{equation}
\hspace{-6em}
 n_{\rm inst}(T,\muq) = \Bigl(\frac{8\pi^2}{g^2}\Bigr)^{2\Nc} \!
  \rme^{-8\pi^2/g(\rho)^2} \rho^{-5}\,\exp\biggl[ -\pi^2\rho^2 T^2
 \Bigl(\frac{2\Nc}{3}+\frac{\Nf}{3}\Bigr) -\Nf\rho^2\muq^2 \biggr] ,
\label{eq:mc-expand}
\end{equation}
in the one-loop calculation with the instanton size whose empirical
value is $\rho\sim 0.3\fm$ \cite{Schafer:1996wv}.  It is thus expected
that the $\ua$-breaking interaction is exponentially suppressed as $T$
and/or $\muq$ grow larger.  If one performs the $\rho$-integration
(which is not IR divergent thanks to the exponential suppression), the
suppression factor would be $T^{-14}$ or $\muq^{-14}$ for $\Nc=\Nf=3$
by the dimensional reason.  It is a non-trivial question whether the
instanton-induced interaction drops by either exponential or power
function \cite{Schafer:2002ty}.  The Columbia plot with an in-medium
reduction of $\ua$ breaking should have a smaller region for the
first-order phase transition.  This has been confirmed at $\muq=0$
\cite{Fukushima:2008wg} in the Polyakov-loop extended chiral
models such as the PNJL model \cite{Fukushima:2003fw,Ratti:2005jh}
and the PQM model \cite{Schaefer:2009ui}.  Consequently the location
of the QCD critical point at finite $\muq$ would be very sensitive to
the strength $c_0$
\cite{Fukushima:2008wg,Fu:2007xc,Schaefer:2008hk}.  The negative
coefficients $c_2$ and $c_4$ in \eref{eq:c24} might be attributed to
the effective $\ua$ restoration at larger $\muq$ \cite{Chen:2009gv};
it has been found that the chiral model analysis with an exponential
ans\"{a}tz, $c_0\propto \rme^{-\Nf\rho^2\muq^2}$, leads to a
scenario similar to the right panel of \fref{fig:curve}.


\subsection{QCD critical point search}
\label{sec:critical-point}
 
In this subsection we summarize physical consequences of the critical
point under the assumption that it exists in the QCD phase diagram.

\begin{figure}
 \begin{center}
 \includegraphics[width=0.4\textwidth]{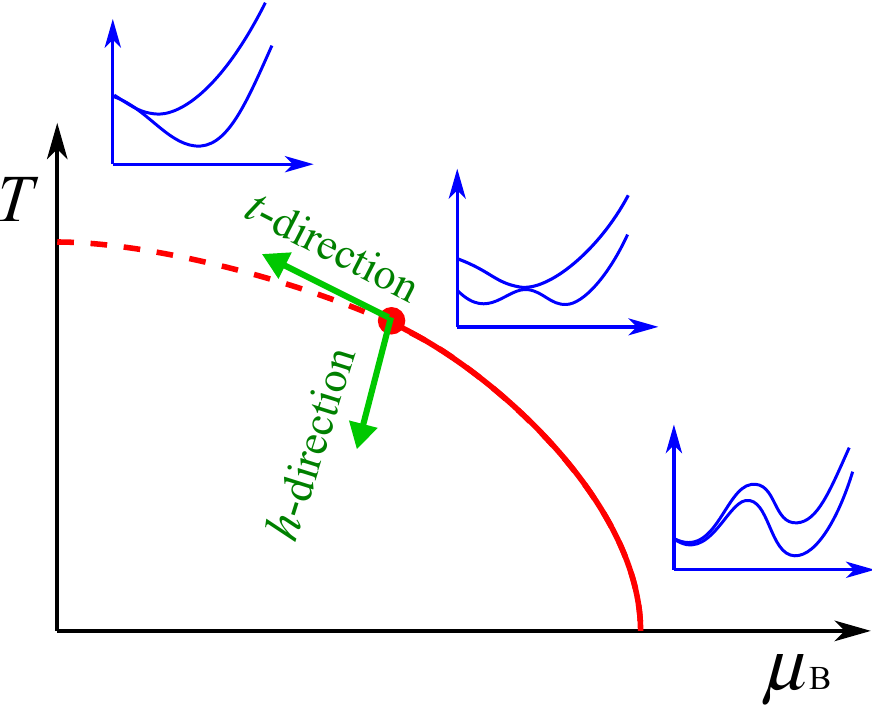}
 \end{center}
 \caption{Schematic illustration of the potential shapes in the
   crossover, critical and first-order regions, respectively, from the
   left to the right.  The $t$-direction at the critical point is
   tangential to the first-order phase boundary.}
 \label{fig:cep}
\end{figure}

\Fref{fig:cep} is a schematic illustration of the shape of the
effective potentials in the crossover and first-order transitions.  If
the critical point is approached along the first-order phase boundary,
the associated potential shape looks like that of the second-order
phase transition around a non-vanishing order parameter.  The critical
properties at the critical point can be mapped to the 3D Ising model
with $\mathrm{Z}(2)$ symmetry with the (reduced) temperature $t$ and
the external field $h$.  On the $\muB$--$T$ plane in the QCD phase
diagram the mapped $t$-direction is tangential to the first-order
phase boundary because $\mathrm{Z}(2)$ symmetry at the critical point
is locally preserved along this direction.  The determination of the
$h$-direction, on the other hand, requires microscopic calculations.


In the vicinity of the critical point along the $t$-direction the
correlation length $\xi$, the chiral condensate $\sigma$ and
 the chiral susceptibility $\chi_\sigma$ have the following
scaling;
\begin{equation}
 \xi \sim t^{-\nu} , \qquad
 \sigma = \langle\bar{\psi}\psi\rangle
  - \langle\bar{\psi}\psi\rangle_0 \sim t^\beta, \qquad
 \chi_\sigma \sim t^{-\gamma} ,
\end{equation}
where $\nu\simeq 0.63$, $\beta\simeq 0.33$ and $\gamma\simeq 1.24$ are
known critical exponents in the 3D Ising model.  Because chiral
symmetry is explicitly broken by $\mq\neq0$, the chiral condensate at
the critical point is non-vanishing and takes a finite
$\langle\bar{\psi}\psi\rangle_0$.  In the same way, along the
$h$-direction, they scale as
\begin{equation}
 \xi \sim h^{-\nu/\beta\delta}, \qquad
 \sigma \sim h^{1/\delta}, \qquad
 \chi_\sigma \sim h^{-1+1/\delta} ,
\end{equation}
where $\delta\simeq 4.8$.  It is also possible to parametrize the
singular part of the equation of state in a similar way as to the 3D
Ising model \cite{Nonaka:2004pg}.

The divergent correlation length $\xi$ can be interpreted physically
as the vanishing screening mass in the $\sigma$-meson channel, which
may have significant consequences in the HIC experiments
\cite{Stephanov:1998dy,Stephanov:1999zu}.  However, one should keep in
mind that the pole mass in the $\sigma$ channel can never be massless
since it is not directly related to $\xi$
\cite{Scavenius:2000qd,Fujii:2003bz}.

In a finite-density medium there is a mixing between the chiral
condensate and the baryon density.  Therefore the divergence in the
chiral susceptibility can be seen also in the baryon number
susceptibility, leading to enhanced fluctuation in the baryon number
$\langle(\delta N)^2\rangle$.  It has been argued that the general
moments (cumulants) become \cite{Stephanov:2008qz};
\begin{equation}
 \langle(\delta N)^k\rangle_{\rm c} \sim \xi^{k(5-\eta)/2-3} ,
\end{equation}
where $\langle\cdots\rangle_{\rm c}$ represents the connected piece of
the correlation function.  Naturally the higher moments have stronger
divergence in the vicinity of the critical point where
$\xi\to\infty$.  On the other hand, they require a large subtraction
like $\langle(\delta N)^4\rangle-3\langle(\delta N)^3\rangle$.  It
would be therefore more difficult to extract them accurately from
HIC experiment data.

Here we emphasize that the soft mode responsible for the critical
property at the critical point is the density fluctuation rather than
the chiral fluctuation.  To see this point it is useful to consider
the following Ginzburg-Landau expansion
\cite{Son:2004iv,Fujii:2004jt};
\begin{eqnarray}
 F[\sigma,\rho] &=& -\frac{\omega^2}{\Gamma}\sigma^2
   -\frac{\rmi\omega}{\lambda\bq^2}\rho^2 + V[\sigma,\rho] \\
 V[\sigma,\rho] &=& a\sigma^2 + b\sigma^4
  + c\sigma^6 - h\sigma + \gamma\sigma^2\rho
  + \frac{1}{2}\rho^2 - j\rho .
\end{eqnarray}
Here $\rho$ represents one of conserved charge densities such as
the baryon density $\nB$ with an appropriate normalization.  The
equilibrium values of $\sigma$ and $\rho$ are fixed by
$\partial V/\partial\sigma=0$ and $\partial V/\partial\rho=0$.  The
dynamics is solved by the kinetic equations of motion
$\partial F/\partial\sigma=0$ and $\partial F/\partial\rho=0$,
which leads to the eigen-frequencies \cite{Fujii:2004jt},
\begin{equation}
 \chi_\sigma^{-1} = \frac{\omega_{\sigma}^2}{\Gamma}
   = \chi_h^{-1} + 4\gamma^2\sigma^2 ,
 \qquad
 \chi_\rho^{-1} = \frac{\rmi\omega_\rho}{\lambda\bq^2}
   = \frac{\chi_h^{-1}}{\chi_h^{-1}+4\gamma^2\sigma^2} ,
\end{equation}
where $\omega_\sigma$ and $\omega_\rho$ are eigen-frequencies which
are identified as the chiral and density modes, respectively.  The
notation $\chi_h$ represents the chiral susceptibility without the
density mixing taken into account, which diverges at the critical
point.  It is apparent from the above expressions that the soft mode
at the QCD critical point is the density fluctuation and the
$\sigma$ mode is a decoupled fast mode.  This also explains in a
natural way why only the screening mass of the $\sigma$ meson becomes
vanishing at the critical point, while the pole mass never does
as mentioned above \cite{Fujii:2003bz}.


\subsection{First-order phase transition at high baryon density}

We now turn to the  chiral phase transition at low temperature
and high baryon density where lattice simulations are not reachable
yet.  In this region, only qualitative analyses based on various
chiral effective theories are available at present.  Many of these
models predict  a first-order transition in cold and dense
matter;  we are going to extract some common feature in these models
based on a quasi-particle picture at $\muq\neq0$.

Let us introduce a vacuum pressure $P_{\rm vac}[\Mq]$, which embodies
the dynamical chiral symmetry breaking, as a function of $\Mq$
(constituent quark mass) as
\begin{equation}
 P_{\rm vac}[\Mq] = -a(M_0^2-\Mq^2)^2 ,
\label{eq:P_chi}
\end{equation}
with a positive curvature parameter $a$.  The maximum of
$P_{\rm vac}[\Mq]$ is realized when $\Mq = \pm M_0 \neq 0$.  (If the
current quark mass $\mq$ is non-zero, a linear term in $\Mq$, which
favours $\Mq >0$ at the maximum of $P_{\rm vac}[\Mq]$, should be
present.)  At zero temperature, as long as $\muq$ is smaller than
$\Mq$, nothing happens and the vacuum remains empty, while $\nB$
starts arising for $\muq>\Mq$.  The pressure from finite $\muq$ is
expressed as
\begin{equation}
\hspace{-3em}
 P_\mu[\Mq] = \int_0^{\muq} d\muq' \;n_{\rm q}(\muq')
 = \frac{\nu}{6\pi^2} \int_0^{\muq} d\muq'\;
   (\muq'^2-\Mq^2)^{3/2}\theta(\muq'^2-\Mq^2) ,
\label{eq:p_mu}
\end{equation}
where $n_{\rm q}(\muq)$ represents the fermion density in the
quasi-particle approximation and
$\nu$=(spin)$\times$(colour)$\times$(flavour) is the fermion
degeneracy.  Because more particles can reside in the Fermi sphere for
smaller mass, $P_\mu[\Mq]$ naturally has a maximum at $\Mq=0$ and goes
to zero at $\Mq=\muq$.

Here $P_{\rm vac}[\Mq]$ and $P_\mu[\Mq]$ have a peak at $\Mq=M_0$ and
$\Mq=0$, respectively.  Assuming that the total pressure is simply
$P[\Mq]=P_{\rm vac}[\Mq]+P_\mu[\Mq]$,  we can see that the
existence of  two separate peaks in $P[\Mq]$ requires
\cite{Fukushima:2008is},
\begin{equation}
 a < \frac{\nu}{16\pi^2}\frac{\muq^2}{M_0^2} 
  \lesssim \frac{\nu}{16\pi^2} \simeq 0.076 ,
\label{eq:curvature}
\end{equation}
for $2$-flavour case with $\nu=12$.  At the first-order critical point
the peak at $\Mq=0$ is just as high as the second peak at
$\Mq\simeq M_0$, which determines the critical
 chemical potential,
\begin{equation}
 a \simeq \frac{\nu}{24\pi^2}\frac{\muc^4}{M_0^4} .
\label{eq:critical}
\end{equation}
Once $a$ satisfies \eref{eq:curvature}, the chiral phase transition
with increasing $\muq$ at $T=0$ should be of first order at
$\muq\simeq\muc$.

The actual value of $a$ is model-dependent:  In the NJL model with two
flavours and in the linear-$\sigma$ model, $a$ is estimated
respectively as
\begin{equation}
 a = \cases{
  \frac{1}{2M_0^2}\biggl( \frac{\nu\Lambda^2}{8\pi^2}
   -\frac{1}{4G_{\rm S}} \biggr) = 0.067 & \mbox{(NJL model)} \cr
  \frac{M_\sigma^2 f_\pi^2}{8M_0^4} = 0.02\sim0.05
   & \mbox{(linear-$\sigma$ model)} \cr},
\end{equation}
where we used the standard NJL parameters; $\Lambda=631\MeV$,
$G_{\rm S}\Lambda^2=2.19$, and the resultant $M_0=336.2\MeV$
\cite{Hatsuda:1994pi}.  The uncertainty in the linear-$\sigma$ model
comes from the choice of the $\sigma$ meson mass.  Then, in both
cases, the estimated $a$ satisfies the inequality \eref{eq:curvature}
implying the first-order phase transition.  The critical chemical
potential deduced from \eref{eq:critical} is, in the NJL model case,
given by $\muc=1.07M_0\simeq 360\MeV$ which is consistent with that
obtained numerically in the NJL model.

Let us now argue that  the first-order transition obtained as
above is rather sensitive to the choice of the model Lagrangian.
Indeed, it has been known that the repulsive contribution to the
pressure of the form $+G_{\rm V}\nq^2$ induced by a
 quark interaction of density-density type can totally wash out
the first-order transition
\cite{Klimt:1990ws,Lutz:1992dv,Kitazawa:2002bc,Sasaki:2006ws,%
Sakai:2008ga,Fukushima:2008is}.  For example, in the NJL model with
the density-density interaction, the condition \eref{eq:curvature} is
changed to
\begin{equation}
 a < \frac{\nu}{16\pi^2}\biggl( 1-\frac{2\nu G_{\rm V}\muq^2}
  {3\pi^2} \biggr) ,
\end{equation}
which implies that the first-order phase transition does not arise for
$G_{\rm V}>0.25G_{\rm S}$
\cite{Kitazawa:2002bc,Sasaki:2006ws,Fukushima:2008is}.


\section{Formation of the diquark condensate}
\label{sec:diquark}

Finding a ground state of quark matter at $T\approx0$ with extremely
large value of $\muq$ is an interesting theoretical challenge.  (We
use $\muq$ instead of $\muB$ throughout this section,  for our
central interest is the quark degrees of freedom).  Let us consider
the Cooper's stability-test in quark matter
\cite{Barrois:1977xd,Bailin:1983bm}.  In the perturbative regime of
QCD the one-gluon exchange potential is proportional to a product of
the quark $\mathrm{SU}(\Nc)$ charges;
\begin{equation}
\hspace{-6em}
 (t^a)_{\alpha \beta}(t^a)_{\alpha' \beta'} 
 = -\frac{\Nc+1}{4\Nc}
     \bigl(\delta_{\alpha \beta}\delta_{\alpha' \beta'}
     -\delta_{\alpha \beta'} \delta_{\alpha' \beta}\bigr)
    +\frac{\Nc-1}{4\Nc}
     \bigl(\delta_{\alpha \beta}\delta_{\alpha' \beta'}
     +\delta_{\alpha \beta'}\delta_{\alpha' \beta}\bigr).
\label{eq:decomposition}
\end{equation}
The first term in the right-hand side of \eref{eq:decomposition} with
negative sign is anti-symmetric under the exchange of colour indices;
$\alpha \leftrightarrow \alpha'$ or $\beta \leftrightarrow \beta'$, so
that a quark pair in the colour anti-triplet channel has attraction.
The second term in the right-hand side of \eref{eq:decomposition} with
positive sign is symmetric under the same exchange of colour indices,
so that a quark pair in the colour sextet channel has repulsion.
Fermi system with two particles attracting each other on a sharp Fermi
sphere has an instability towards the formation of Cooper pairs.
Therefore, normal quark matter inevitably becomes
colour-superconducting (CSC) phase with diquark condensate at
asymptotic high density and at sufficiently low temperature
\cite{Barrois:1977xd,Bailin:1983bm,Iwasaki:1994ij,Alford:1997zt,%
Rapp:1997zu}.

Since quarks carry not only spin but also colour and flavour, various
pairing patterns are possible.  Hereafter we use the following
notation; the colour indices $\alpha$, $\beta$, $\gamma$ run from $1$
to $3$ meaning $r$ (red), $g$ (green) and $b$ (blue) in order, and in
the same way the flavour indices $i$, $j$, $k$ run from $1$ to $3$
meaning $u$ (up), $d$ (down) and $s$ (strange) in order.  We will
focus our attention to quark matter below the charm threshold,
i.e.\ $\muq < m_{\rm charm}$, so that we do not take into account
heavy flavours ($c$, $b$ and $t$).


\paragraph{Spin-zero condensate{\rm :}}
The quark pairing with zero total spin is characterized by the
following order parameter with $3\times 3$ matrix structure
\cite{Alford:1998mk};
\begin{equation}
 (d^{\dagger})_{\alpha i} \sim
  \epsilon_{\alpha\beta\gamma}\, \epsilon_{ijk}
  \langle \psi^{\rm t}_{\beta j} \,C \gamma^5\,
  \psi_{\gamma k} \rangle .
\label{eq:spin0-cond}
\end{equation}
The quark pair in this case is in the colour anti-symmetric
(anti-triplet), flavour anti-symmetric (anti-triplet) and spin
anti-symmetric channel to satisfy the Fermi statistics.  The charge
conjugation matrix $C=\rmi\gamma^2\gamma^0$ together with $\gamma^5$
makes the above condensate a Lorentz scalar.  (The effect of
instantons and also quark masses favours the scalar condensate instead
of the pseudo-scalar condensate.)  If the masses of $u$, $d$ and $s$
quarks are all degenerate, there is an exact flavour
$\mathrm{SU}_{\rm V}(3)$ symmetry.  Then, one can always diagonalize
the order parameter by the bi-unitary rotation in colour and flavour
space to obtain $(d)_{i \alpha}=\delta_{i \alpha} \Delta_i$.  We will
adopt this as an ans\"{a}tz even when $u$, $d$ and $s$ quarks are not
degenerate, which is in practise a good approximation
\cite{Alford:1999pa,Ruester:2004eg}.

Let us introduce a notation $\Dds=\Delta_1$, $\Dsu=\Delta_2$,
$\Dud=\Delta_3$ to indicate which quarks are involved in the Cooper
pairing.  If all the gaps are non-vanishing ($\Dud\neq0$, $\Dds\neq0$
and $\Dsu\neq0$),  such a state is called the colour-flavour locked
(CFL) phase \cite{Alford:1998mk} because the colour and flavour
degrees of freedom are entangled with each other.  For $\Dds =0$ with
other components non-vanishing, it is called the uSC phase since both
$\Dud$ and $\Dsu$ contain the $u$ quark.  The dSC and sSC phases are
defined in the same way
\cite{Iida:2003cc,Ruester:2004eg,Fukushima:2004zq}.  If only $\Dud$ is
non-vanishing, it is called the 2SC (two-flavour superconducting)
phase.  In case that we refer to similar phases with other flavour
combination, we write the 2SCds, 2SCsu or 2SCud($=$2SC) phase
\cite{Alford:1999pa,Alford:2004nf}.  When all the pairing gaps are
absent, the system is in the state of normal quark matter (NQM).  We
summarize these abbreviations in \tref{tab:class}.


\begin{table}
 \begin{center}
 \begin{tabular}{c|cccccccc}
  \hline
  Pairing & CFL & uSC & dSC & sSC & 2SC & 2SCds & 2SCsu & NQM\\
  \hline
  $\Delta_{ud}$ & $\bigcirc$ & $\bigcirc$ & $\bigcirc$ & $\times$
   & $\bigcirc$ & $\times$   & $\times$ & $\times$ \\
  $\Delta_{ds}$ & $\bigcirc$ & $\times$   & $\bigcirc$ & $\bigcirc$
   & $\times$   & $\bigcirc$ & $\times$ & $\times$ \\
  $\Delta_{su}$ & $\bigcirc$ & $\bigcirc$ & $\times$   & $\bigcirc$
   & $\times$   & $\times$   & $\bigcirc$ & $\times$ \\
  \hline
 \end{tabular}
 \end{center}
 \caption{Classification of the colour-superconducting phases.}
 \label{tab:class}
\end{table}


Let us now consider the symmetry breaking patterns ($\calG\to\calH$)
in the CFL and 2SC phases.
\footnote{Similar comment on the quotient groups ($\calG'$ and
  $\calH'$) as given to \eref{eq:patternG} and \eref{eq:patternH} is
  applied here.}
In the CFL phase for massless 3-flavours, one finds the following
pattern \cite{Alford:1998mk,Alford:2007xm},
\begin{eqnarray}
\label{eq:calG} 
 \calG &=& \mathrm{SU}(3)_{\rm C} \times
  \underbrace{\mathrm{SU}(3)_{\rm L} \times
  \mathrm{SU}(3)_{\rm R}}_{\displaystyle \supset \mathrm{U}(1)_{\rm em}}
  \times \mathrm{U}(1)_{\rm B}\times \mathrm{Z}(6)_{\rm A}, \\
\label{eq:calH}
 \calH &=& \underbrace{\mathrm{SU}(3)_{\rm C+L+R}}_{\displaystyle
  \supset \mathrm{U}(1)_{\rm em+C} } \times \mathrm{Z}(2)_{\rm V} .
\end{eqnarray}
Here the baryon number symmetry is broken to leave its discrete
subgroup $\mathrm{Z}(2)_{\rm V}$ which corresponds to a reflection
$\psi \rightarrow -\psi$.  Also, (global) colour symmetry and
chiral symmetry are broken simultaneously to leave their diagonal
subgroup $\mathrm{SU}(3)_{\rm C+L+R}$ intact.  Note that the
electromagnetic symmetry $\mathrm{U}(1)_{\rm em}$ associated with the
electric charge $Q=\diag(2/3,-1/3,-1/3)$ happens to be a subgroup of
the vector symmetry $\mathrm{SU}(3)_{\rm L+R}$.  In the CFL phase
$\mathrm{U}(1)_{\rm em}$ survives as a \textit{modified} symmetry,
$\mathrm{U}(1)_{\rm em+C}$, which contains simultaneous
electromagnetic and colour rotation.  As a result, seven gluons and
one gluon-photon mixture acquire finite Meissner mass by the
Anderson-Higgs mechanism, while the other photon-gluon mixture
remains massless.

The 2SC phase is quite different from the CFL phase in the sense that
only $u$ and $d$ quarks participate in the pairing with a chiral
flavour-singlet combination.  The symmetry breaking pattern in this
case for massless 2-flavours with infinitely heavy strange quarks
reads \cite{Alford:2007xm},
\begin{eqnarray}
\label{eq:calG2} 
 \calG &=& \mathrm{SU}(3)_{\rm C} \times
  \underbrace{\mathrm{SU}(2)_{\rm L} \times
  \mathrm{SU}(2)_{\rm R} \times \mathrm{U}(1)_{\rm B}
  }_{\displaystyle \supset \mathrm{U}(1)_{\rm em}}
  \times \mathrm{Z}(4)_{\rm A}, \\
\label{eq:calH2} 
 \calH &=& \mathrm{SU}(2)_{\rm C} \times
  \underbrace{\mathrm{SU}(2)_{\rm L} \times
  \mathrm{SU}(2)_{\rm R} \times
  \mathrm{U}(1)_{\rm C+B}
  }_{\displaystyle \supset \mathrm{U}(1)_{\rm em+C}} .
\end{eqnarray} 
Note that chiral symmetry remains intact while colour symmetry is
broken, so that five gluons (4th to 8th) receive finite Meissner
mass.  In the 2SC phase, the baryon-number symmetry
$\mathrm{U}(1)_{\rm B}$ survives as a \textit{modified} symmetry,
$\mathrm{U}(1)_{\rm C+B}$, which contains simultaneous electromagnetic
and 8-th colour rotation.  The global electromagnetism which is
originally a combination of the baryon number symmetry and the isospin
symmetry, survives also as a \textit{modified} symmetry,
$\mathrm{U}(1)_{\rm em+C}$.  Therefore, the 2SC phase is neither a
superfluid nor an electromagnetic superconductor.  The modified
symmetries discussed above remain unbroken in the uSC and dSC phases
as well.

The CFL phase is reminiscent of what is called the B phase of
superfluid $^3$He in which fermionic $^3$He atoms have a spin-triplet
($S=1$) and $p$-wave ($L=1$) pairing.  The B phase is a state in which
the orbital angular momentum $\boldsymbol{L}$ and the spin
$\boldsymbol{S}$ are locked so that the symmetry breaking pattern is
$\mathrm{SO}(3)_{\rm S} \times \mathrm{SO}(3)_{\rm L} \times
\mathrm{U}(1)_\phi \,\to\, \mathrm{SO}(3)_{\rm S+L}$, where
$\mathrm{U}(1)_\phi$ is a symmetry corresponding to
$\mathrm{U}(1)_{\rm B}$ in quark matter.


\paragraph{Spin-one condensate{\rm :}}
There is also a possibility of flavour symmetric Cooper pairs such as
the pairing within the same flavour \cite{Iwasaki:1994ij}.  In this
case $s$-wave Cooper pair must be spin triplet ($J=L+S=0+1=1$) to
satisfy the Fermi statistics.

Various forms of the spin-one CSC states have been studied so far
\cite{Alford:2004nf,Schmitt:2002sc,Schmitt:2003xq,Schmitt:2004et}.
Since the Cooper pair is triplet in both colour and spin, the order
parameter becomes a complex $3\times3$ matrix, which is similar to the
situation in the CFL phase where the Cooper pair is triplet in both
colour and flavour.  The explicit form would be
\begin{equation}
 (\widetilde{d}^{\dagger})_{\alpha i} \sim  \epsilon_{\alpha\beta\gamma}
  \langle\psi^{\rm t}_\beta C\gamma^i \psi_\gamma \rangle ,
\end{equation}
where $i$ refers to not the flavour but the Lorentz index.  The
colour-spin locked (CSL) ans\"{a}tz,
$(\widetilde{d})_{i\alpha} \propto \delta_{i \alpha}$, leads to a
symmetry breaking pattern which is similar to the CFL phase in CSC and
to the B phase in superfluid $^3$He;
\begin{equation}
 \mathrm{SU}(3)_{\rm C} \times \mathrm{SU}(2)_{\rm J}
  \times \mathrm{U}(1)_{\rm em} \;\to\;
  \mathrm{SU}(2)_{\rm C+J} .
\end{equation}
On the other hand, if we take an ans\"{a}tz, $(\widetilde{d})_{\alpha i}
\propto \delta_{\alpha 3}(\delta_{i1}+\rmi \delta_{i2})$, we have a
symmetry breaking pattern,
\begin{equation}
 \mathrm{SU}(3)_{\rm C} \times \mathrm{SU}(2)_{\rm J}
  \times \mathrm{U}(1)_{\rm em} \;\to\;
  \mathrm{SU}(2)_{\rm C} \times \mathrm{U}_{\rm J+em} \times
  \mathrm{U}(1)_{\rm em+C} ,
\end{equation}
which is analogous to the A phase of superfluid $^3$He in which the
symmetry breaking patter is; $\mathrm{SO}(3)_{\rm S} \times
\mathrm{SO}(3)_{\rm L} \times \mathrm{U}(1)_\phi \;\to\;
\mathrm{U}(1)_{\rm S_z} \times \mathrm{U}(1)_{\rm L_z+\phi}$.  Other
pairing patterns such as the polar phase, the planar phase and so on
have been also investigated \cite{Schmitt:2004et}.


\paragraph{Weak-coupling results{\rm :}}
In the asymptotically high-density region, the running coupling
constant of the strong interaction becomes small enough to justify the
weak-coupling calculations, so that the diquark condensate as well as
the gap energy of quarks can be estimated from the first-principle QCD
calculation.  The driving force of the diquark condensate is the
interaction between quarks due to gluon exchange.  The chromoelectric
part of the interaction in high-density quark matter is screened by
the Debye mass $m_{_{\rm D}} = \sqrt{\Nf/(2 \pi^2)}\, g\muq$, while
the choromomagnetic part of the interaction is screened only
dynamically by Landau damping \cite{Son:1998uk}.  Therefore, instead
of having the standard BCS form,
$\Delta \sim \muq\, \exp(-{\rm const.}/g^2)$, the gap energy at the
Fermi surface is parametrically enhanced due to large forward
scattering between quarks \cite{Son:1998uk,Hong:1999fh,Schafer:1999jg,%
Pisarski:1999tv,Schmitt:2002sc};
\begin{equation}
\hspace{-2em}
 \Delta_{\rm F} = 512\pi^4 \biggl(\frac{2}{\Nf}\biggr)^{5/2} \!
  \rme^{-(\pi^2+4)/8}\, (\lambda_1^{a_1}\lambda_2^{a_2})^{-1/2}\; g^{-5}
  \muq\, \exp\biggl(-\frac{3\pi^2}{\sqrt{2}g}\biggr) ,
\label{eq:gap-energy}
\end{equation}
where $\lambda_1=a_1=1$ and $\lambda_2=a_2=0$ for the 2SC phase, while
$\lambda_1=4$ and $\lambda_2=1$ with $a_1=1/3$ and $a_2=2/3$ in the
CFL phase.  In the exponent of \eref{eq:gap-energy}, unusual
dependence on the coupling $g^{-1}$ instead of the standard dependence
$g^{-2}$ enhances the gap substantially in the weak coupling regime
both for 2SC and CFL phases.  We see that $\Delta^{\rm 2SC}$ in the
2SC phase is greater than that in the CFL phase as
$\Delta^{\rm 2SC}=2^{1/3}\Delta^{\rm CFL}$, which holds not only in
the weak-coupling QCD calculation but also in effective model
calculations.  Similarly to the BCS theory, the gap energy at zero
temperature and the critical temperature are related to each other in
the mean-field approximation \cite{Schmitt:2002sc},
\begin{equation}
 \Tc = (\lambda_1^{a_1}\lambda_2^{a_2})^{1/2} \;
  \frac{\rme^{\gamma}}{\pi}\Delta_{\rm F} .
\end{equation}
Note that the factor $(\lambda_1^{a_1}\lambda_2^{a_2})$ in the
right-hand side is cancelled by the same factor in
\eref{eq:gap-energy}, so that $\Delta^{\rm 2SC}>\Delta^{\rm CFL}$ does
not necessary imply $\Tc^{\rm 2SC}>\Tc^{\rm CFL}$.


\subsection{Neutrality in electric and colour charges}

For the bulk system to be stable, the total electric and colour
charges must be zero due to the Gauss law
\cite{Alford:2002kj,Steiner:2002gx,Gerhold:2003js}.  In some effective
models which do not have gauge fields as explicit degrees of freedom,
the charge neutrality must be imposed by  electric and colour
chemical potentials,
$\mu_{\alpha i} = \muq + \mu_{Q}(Q)_{ii} + \mu_3 (T_3)_{\alpha\alpha}
 + \mu_8 ({\scriptstyle \frac{2}{\sqrt{3}}}T_8)_{\alpha\alpha}$. 
Their explicit forms read
\begin{equation}
\hspace{-3em}
 \begin{array}{lp{1em}l}
  \mu_{ru} = \muq -\frac{2}{3}\mue + \frac{1}{2}\mu_3 +
   \frac{1}{3}\mu_8 , &&
  \mu_{gd} = \muq +\frac{1}{3}\mue - \frac{1}{2}\mu_3 +
   \frac{1}{3}\mu_8 , \\[3pt]
  \mu_{bs} = \muq +\frac{1}{3}\mue - \frac{2}{3}\mu_8 , &&
  \mu_{rd} = \muq +\frac{1}{3}\mue + \frac{1}{2}\mu_3 +
   \frac{1}{3}\mu_8 , \\[3pt]
  \mu_{gu} = \muq -\frac{2}{3}\mue - \frac{1}{2}\mu_3 +
   \frac{1}{3}\mu_8 , &&
  \mu_{rs} = \muq +\frac{1}{3}\mue + \frac{1}{2}\mu_3 +
   \frac{1}{3}\mu_8 , \\[3pt]
  \mu_{bu} = \muq -\frac{2}{3}\mue - \frac{2}{3}\mu_8 , &&
  \mu_{gs} = \muq +\frac{1}{3}\mue - \frac{1}{2}\mu_3 +
   \frac{1}{3}\mu_8 , \\[3pt]
  \mu_{bd} = \muq +\frac{1}{3}\mue - \frac{2}{3}\mu_8 ,
 \end{array}
\end{equation}
where the electron chemical potential $\mue$ satisfies
$\mue=-\mu_Q$ due to $\beta$-equilibrium.  These constraints add some
varieties in the physics of CSC because one needs to consider the
Cooper pairing between quarks with mismatched Fermi surfaces.

\begin{figure}
 \begin{center}
 \includegraphics[width=0.25\textwidth]{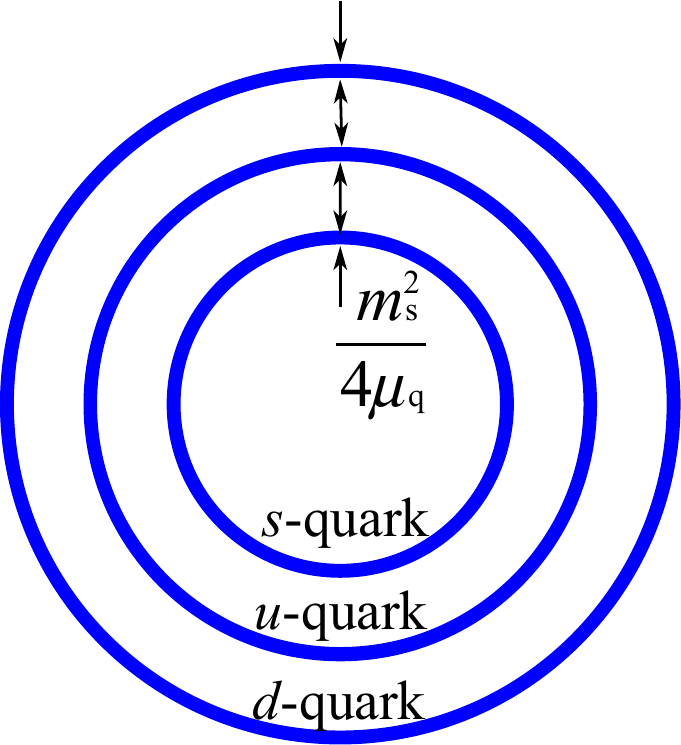}
 \end{center}
 \caption{Fermi surface mismatch in unpaired normal quark matter.}
 \label{fig:fermisphere}
\end{figure}

For unpaired quark matter with $\mud=0$ and $\mst\neq0$, the number of
strange quarks is less than that of other quarks.  There arise
electrons in the system to maintain charge neutrality accordingly.
The neutrality conditions obtained by the variation of the free
energy,
$\partial \Omega_{\rm unpaired}/\partial \mue =
\partial \Omega_{\rm unpaired}/\partial \mu_3 =
\partial \Omega_{\rm unpaired}/\partial \mu_8 =0$, lead to
$\mu_3=\mu_8=0$ and $\mue=\mst^2/(4\muq)$.  With the free dispersion
relation at large chemical potential,
$\epsilon_{\rm ud}(p) = |{\bp}|$ and
$\epsilon_{\rm s}(p) = \sqrt{|{\bp}|^2 + \mst^2} \simeq
 |{\bp}| +\mst^2/(2\muq)$, the Fermi momenta have mismatches as
\begin{equation}
 p_{\rm F}^u = \muq - \frac{\mst^2}{6\muq} ,\ \quad
 p_{\rm F}^d = \muq + \frac{\mst^2}{12\muq} ,\ \quad
 p_{\rm F}^s = \muq - \frac{5\mst^2}{12\muq} .
\end{equation}
This is illustrated in \fref{fig:fermisphere}.


At high density where the expansions in terms of $\mst/\muq$ and
$\Delta/\muq$ are valid, model-independent conclusions can be drawn
for the CSC quark matter at $T=0$ \cite{Alford:2002kj,Steiner:2002gx}.
In the CFL phase the modified electric charge $\Qtilde$ associated
with the symmetry $\mathrm{U}(1)_{\rm em+C}$ is conserved.  Also, all
the Cooper pairs have $\Qtilde=0$, so that the CFL quark matter at
$T=0$ is a $\Qtilde$-insulator \cite{Rajagopal:2000ff}.  Thus, the CFL
free energy $\Omega_{\rm CFL}$ is independent of the corresponding
chemical potential except for a small contribution proportional to
$\mue^4$ from $\Qtilde$-violating electrons.  Under the neutrality
conditions, $\partial \Omega_{\rm CFL}/\partial \mue =
\partial \Omega_{\rm CFL}/\partial \mu_3 =
\partial \Omega_{\rm CFL}/\partial \mu_8 =0$, we obtain
\begin{equation}
 \mu_3 = \mue ,\qquad
 \mu_8 = \frac{1}{2}\mue -\frac{\mst^2}{2\muq} , \qquad
 \mue=0 .
\end{equation}
It is an important property that the CFL phase
($\Dud\simeq\Dds\simeq\Dsu$ with $\mue=0$) is rigid against the Fermi
surface mismatch \cite{Rajagopal:2000ff} as long as the CFL phase is
in the stable region; $\Delta > \mst^2/(2\muq)$.  We will discuss
later what happens once this stability condition is violated in
\sref{sec:chromomagnetic}.

In the 2SC phase the colour and charge neutralities lead to
\begin{eqnarray}
 \mu_3=\mu_8=0, \qquad \mue=\frac{\mst^2}{2\muq},
\end{eqnarray}
so that electrons are required to be present with even larger chemical
potential than the normal quark matter.  Also, the 2SC phase contains
ungapped $\Qtilde$-carrying modes, and is therefore a
$\Qtilde$-conductor.


\subsection{Ginzburg-Landau approach}

To identify the phase structure of CSC near second-order or weak
first-order transitions at finite $T$,  the Ginzburg-Landau
approach is quite useful \cite{Bailin:1983bm,Iida:2000ha}.  If we
assume that the gap energy is small compared to the critical
temperature, the Ginzburg-Landau free energy with a power series
expansion in terms of $(d)_{i \alpha}$ up to the quartic order reads
\cite{Iida:2003cc},
\begin{eqnarray}
 \Omega &= \alpha \tr d d^\dagger +\beta_1 \bigl( \tr d d^\dagger\bigr)^2
  +\beta_2 \tr(d d^\dagger)^2 . \nonumber\\
 &\quad + \epsilon \sum_\alpha
  \bigl( |d_{\alpha u}|^2+|d_{\alpha d}|^2 \bigr)
  + \frac{\eta}{3} \sum_\alpha \bigl(|d_{\alpha d}|^2
  + |d_{\alpha s}|^2 - 2|d_{\alpha u}|^2 \bigr) ,
\label{eq:melting}
\end{eqnarray}
where tr is taken in flavour space.  For $\mst=\mue=0$ the transition
from the CFL phase to normal quark matter is driven by the parameter
$\alpha$ changing the sign from negative to positive.  The fourth term
with $\epsilon$ takes care of the asymmetry between ($u$, $d$) and $s$
introduced by $\mst\neq0$, while the fifth term with $\eta$ represents
the asymmetry between $u$ and ($d$, $s$) from the charge neutrality
effect by $\mue\neq0$.  The effect of colour neutrality is negligible
near $\Tc$.  Under the diagonal ans\"{a}tz on the gap matrix, the free
energy simplifies into
\begin{eqnarray}
 \Omega &= \alpha'\bigl( \Dud^2+\Dds^2+\Dsu^2 \bigr) - \epsilon\,\Dud^2
  -\eta\,\Dds^2 \nonumber\\
 &\qquad + \beta_1 \bigl( \Dud^2+\Dds^2+\Dsu^2 \bigr)^2
  + \beta_2 \bigl( \Dud^4+\Dds^4+\Dsu^4 \bigr)
\end{eqnarray}
with $\alpha'=\alpha+\epsilon+\eta/3$.  In the weak-coupling analysis
one can show that $\beta_1 = \beta_2 = \beta$ and
$\alpha = \alpha_0 (T-\Tc)/\Tc$, where $\alpha_0 = 2\muq^2/\pi^2$ and
$\Tc$ is the critical temperature defined for $\epsilon=\eta=0$.   In
the weak-coupling limit of QCD with $\mst \ll \muq$, it is shown that
$\epsilon \simeq 2\eta \simeq 2\alpha_0\delta$ with a parameter
$\delta \simeq (\mst^2/8\muq^2) \ln(\muq/\Tc)$.  Then
the gap energies near $\Tc$ behave as
\begin{equation}
\hspace{-7em}
 \Dud^2(T) = \frac{\alpha_0}{8\beta}\biggl( \frac{\Tc-T}{\Tc}
  + \frac{8}{3}\delta \biggr) ,\quad
 \Dds^2(T) = \Dud^2 - \frac{\alpha_0 \delta}{2\beta} ,\quad
 \Dsu^2(T) = \Dud^2 - \frac{\alpha_0 \delta}{\beta} .
\label{eq:ordering}
\end{equation}
The inequalities, $\Dud(T)>\Dds(T)>\Dsu(T)$, imply sequential
melting of CSC as $T$ increases at high baryon density, i.e.\
CFL $\rightarrow$ dSC $\rightarrow $ 2SC $\rightarrow$ NQM.\ \
This scenario of melting pattern has been confirmed by explicit
calculations in the NJL model \cite{Fukushima:2004zq}.


\begin{figure}
 \begin{center}
 \includegraphics[width=0.4\textwidth]{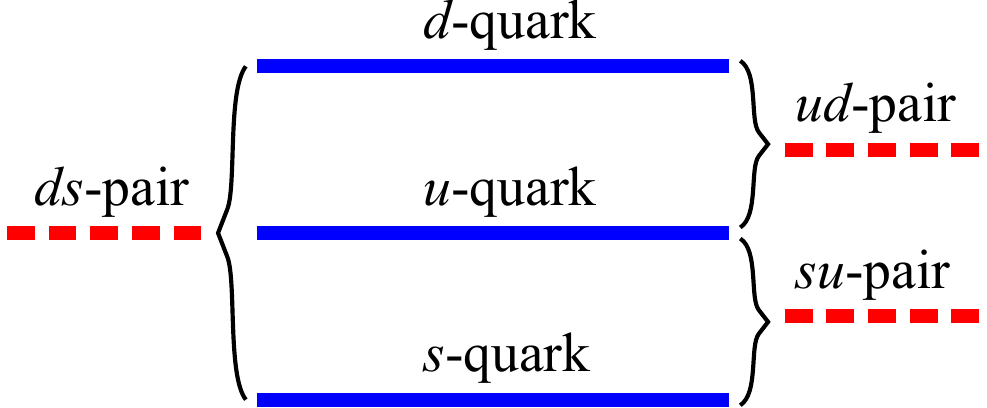}
 \end{center}
 \caption{Ordering of the averaged Fermi surfaces for respective
   Cooper pairs.}
 \label{fig:fermisphere-gl}
\end{figure}


The ordering $\Dud>\Dds>\Dsu$ in \eref{eq:ordering} can be understood
in an intuitive manner.  Because $T$ is close to $\Tc$, the Fermi
surface is blurred and the mismatch of the Fermi momentum no longer
imposes a pressure to take the pairing apart.  The magnitude of the
gap energy is then determined by the density of states.  For example,
the magnitude of the density of states for the pairing $\Dud$ is
related to the size of the ``averaged" Fermi surface over $u$ and
$d$ quarks.  The Fermi surface ordering as shown in
\fref{fig:fermisphere} gives rise to the ordering of the averaged
Fermi surfaces as shown in \fref{fig:fermisphere-gl}, which explains
the ordering $\Dud>\Dds>\Dsu$ near $\Tc$.
 
In some model calculations in the literature, uSC instead of dSC
appears even for extremely  large $\muq$.  This is caused by an
artifact of the UV cutoff $\Lambda$ in the model:  If $\Lambda$ is not
chosen to be sufficiently large, $\epsilon$ and $\eta$ suffer from
significant cutoff-artifact, so that $\eta$ may even  have an
opposite sign from the QCD prediction \cite{Fukushima:2004zq}.  On the
other hand, as $\muq$ becomes small and approaches to $\mst$, the
expansion in terms of $\mst/\muq$ adopted  for the computation of
$\epsilon$ and $\eta$ in \eref{eq:melting} is no longer valid, and the
other melting pattern, CFL $\rightarrow$ uSC $\rightarrow $ 2SC
$\rightarrow$ NQM, indeed becomes possible \cite{Fukushima:2004zq}.


\subsection{Quark-hadron continuity and $\ua$ anomaly}
\label{sec:continuity}

The Ginzburg-Landau approach can be extended to incorporate the
competition between the diquark condensate and the chiral condensate
\cite{Hatsuda:2006ps}.  In this case it is necessary to distinguish
the right-handed and left-handed diquark condensates $\dL$ and $\dR$
as given in \eref{eq:LR-d} to construct properly the Ginzburg-Landau
free energy with the symmetry $\calG$ in \eref{eq:calG}.  The basic
transformation properties of the fields $\Phi$, $\dL$ and $\dR$ are
\begin{eqnarray}
 \Phi \rightarrow  V_{\rm L}^{\phantom{\dagger}}
                   \Phi V_{\rm R}^{\dagger}, \ \ \ 
 \dL  \rightarrow  V_{\rm L}^{\phantom{\dagger}}
                   \dL  V_{\rm C}^{\rm t}, \ \ \ 
 \dR  \rightarrow  V_{\rm R}^{\phantom{\dagger}}
                   \dR  V_{\rm C}^{\rm t}, 
 \end{eqnarray}
where $V_{\rm L}$, $V_{\rm R}$ and $V_{\rm C}$ correspond to
$\mathrm{U}(3)_{\rm L}$, $\mathrm{U}(3)_{\rm R}$ and
$\mathrm{SU}(3)_{\rm C}$ rotations, respectively.

In a simple case where all $u$, $d$ and $s$ quarks are massless, the
chiral part has a standard expansion in the same form as
\eref{eq:chiral-scalar};
\begin{equation}
\hspace{-2em}
 \Omega_\Phi = \frac{a_0}{2}\tr \Phi^\dagger\Phi
  +\frac{b_1}{4!}\bigl(\tr\Phi^\dagger\Phi\bigr)^2
  +\frac{b_2}{4!}\tr(\Phi^\dagger\Phi)^2
  - \frac{c_0}{2}\bigl(\det\Phi + \det\Phi^\dagger\bigr) .
\label{eq:omega_phi}
\end{equation}
The term with the coefficient $c_0$ originates from the axial
anomaly.

The diquark free energy up to the quartic order reads
\begin{eqnarray}
 \Omega_d &=& \alpha \tr\bigl( \dL\dLd + \dR\dRd \bigr)
  + \beta_1 \Bigl[ \bigl(\tr \dL\dLd\bigr)^2
   +\bigl(\tr \dR\dRd\bigr)^2 \Bigr] \nonumber\\
  && + \beta_2 \Bigl[ \tr\bigl(\dL\dLd\bigr)^2
  + \tr\bigl(\dR\dRd\bigr)^2 \Bigr] \nonumber\\
  && + \beta_3 \tr\bigl( \dR\dLd \dL\dRd \bigr)
  + \beta_4 \tr \dL\dLd \tr \dR\dRd .
\label{eq:omega_d}
\end{eqnarray}
The transition from the CFL phase to normal quark matter is driven
by the parameter $\alpha$ changing the sign from negative to
positive.  Unlike $\det\Phi$ in \eref{eq:omega_phi}, terms such as 
$\det \dL$ and $\det \dR$ are not allowed in \eref{eq:omega_d} since
they break $\mathrm{U}(1)_{\rm B}$.

The coupling between the diquark and the chiral condensates has the
following general form up to the quartic order;
\begin{eqnarray}
\hspace{-2em}
 \Omega_{\Phi d} &=& \gamma_1 \tr\bigl( \dR\dLd\Phi
  + \dL\dRd\Phi^\dagger \bigr) \nonumber\\
 && + \lambda_1 \tr\bigl( \dL\dLd\Phi\Phi^\dagger
  + \dR\dRd\Phi^\dagger\Phi \bigr)
  + \lambda_2 \tr\bigl(\dL\dLd + \dR\dRd)
   \tr\Phi^\dagger\Phi \nonumber\\
 && + \lambda_3 \bigl( \det\Phi \tr \dL\dRd\Phi^{-1}
  + \det\Phi^\dagger \tr \dR\dLd\Phi^{\rm t} \bigr) .
\end{eqnarray}
The term with the coefficient $\gamma_1$ originates from the axial
anomaly.

In the massless $3$-flavour limit, one may assume
$\Phi=\diag(\sigma,\sigma,\sigma)$ and
$\dL=-\dR=\diag(\Delta,\Delta,\Delta)$ and then the sum of above three
pieces amounts to
\begin{equation}
\hspace{-2em}
 \Omega_{\rm 3F} = \biggl( \frac{a}{2}\sigma^2 - \frac{c}{3}\sigma^3
  +\frac{b}{4}\sigma^4 \biggr) + \biggl( \frac{\alpha}{2}\Delta^2
  + \frac{\beta}{4}\Delta^4 \biggr) -\gamma\, \Delta^2\sigma
  + \lambda\, \Delta^2\sigma^2 .
\label{eq:3F-Omega}
\end{equation}
Here the $\sigma^3$ and the $\Delta^2\sigma$ terms originate from the
axial anomaly.  From the Fierz transform of the anomaly-induced KMT
interaction in the quark level, the coefficients $c$ and $\gamma$ turn
out to have the same sign \cite{Hatsuda:2006ps}.  Furthermore, the
sign of $c$ and $\gamma$ can be taken to be positive without loss of
generality because of the relation,
$\Omega_{\rm 3F}(\sigma,\Delta; c,\gamma)
 = \Omega_{\rm 3F}(-\sigma,\Delta; -c,-\gamma)$, which is a situation
analougous to the quark mass term.

\begin{figure}
 \begin{center}
 \includegraphics[width=0.55\textwidth]{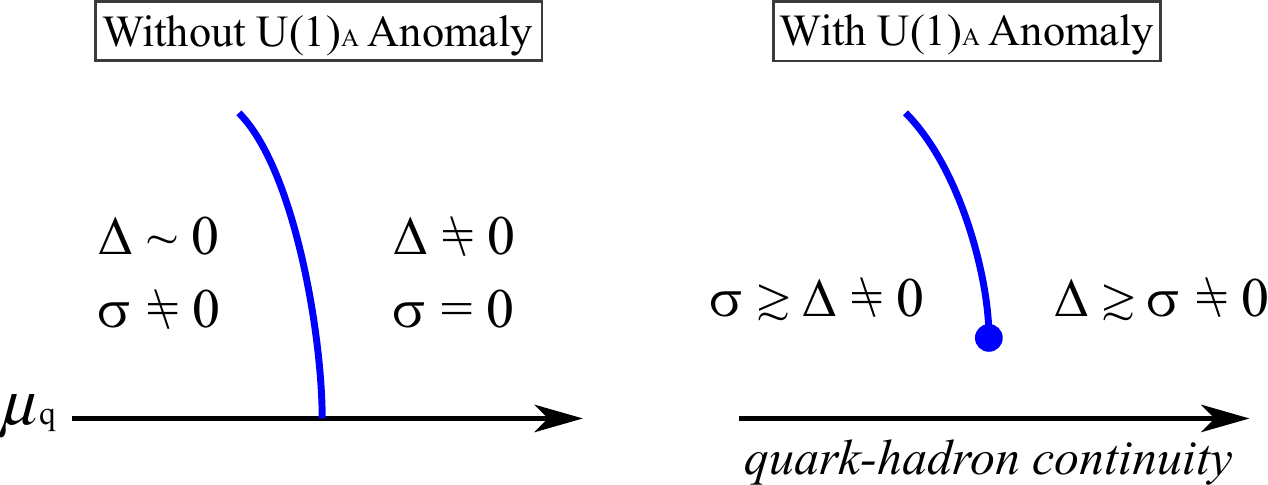}
 \end{center}
 \caption{Schematic figure on the realization of the quark-hadron
   continuity induced by the $\ua$ anomaly.  (Left) First-order phase
   boundary near $T=0$ without the anomaly-induced $\sigma\Delta^2$
   term.  (Right) Smooth crossover and the appearance of a critical
   point due to a large $\sigma\Delta^2$ term with $\gamma >0$ in
   \eref{eq:3F-Omega}.}
 \label{fig:another}
\end{figure}

It should be noted that the $\Delta^2\sigma$ term with positive
coefficient $\gamma$ favours the coexistence of the chiral and diquark
condensates.  Also, this term is linear in $\sigma$ and behaves as if
it is an explicit chiral symmetry breaking term.   In contrast,
$\Delta^2\sigma^2$ term with positive coefficient $\lambda$ (the
positive sign is supported by the weak-coupling calculation and in the
NJL model) disfavours the coexistence.  Therefore, if the effect of
$\gamma$ is sufficiently strong and $\Delta$ is sufficiently large,
the first-order phase boundary, which normally separates the chiral
symmetry breaking phase ($\sigma \neq 0$) and the CFL phase
($\sigma=0$ and $\Delta \neq 0$), can be smeared out.  In such a case
the two phases are connected smoothly  to each other and a critical
point associated with this crossover appears as shown in
\fref{fig:another}.


The smooth crossover of the CFL phase and the Nambu-Goldstone
(hadronic) phase may have  a close connection to the idea of
\textit{quark-hadron continuity} \cite{Alford:1999pa,Schafer:1998ef}.
Since the axial anomaly tends to enhance (reduce) the first-order
transition through the term proportional to $c$ ($\gamma$) in
\eref{eq:3F-Omega}, it is a dynamical issue whether the phase diagram
as sketched in \fref{fig:another} is realized or not in the real world
\cite{Yamamoto:2007ah,Yamamoto:2008zw,Abuki:2010jq}.


\subsection{Collective excitations}

An interesting and related question associated with the $\ua$ anomaly
is the fate of collective excitations.  Using the chiral effective
Lagrangian approach \cite{Casalbuoni:1999wu}, one finds the dispersion
relations for the pions and kaons in the CFL phase with $\mst\neq0$
and $\mue\neq0$  \cite{Son:1999cm,Son:2000tu},
\begin{eqnarray}
 \epsilon_{\pi^\pm}(p)
  &= \pm\mue + \sqrt{v^2 p^2 + M_{\pi^\pm}^2} ,\nonumber\\
 \epsilon_{K^\pm}(p)
  &= \pm\mue \mp\frac{\mst^2}{2\muq} + \sqrt{v^2 p^2 + M_{K^\pm}^2} ,
  \nonumber\\
 \epsilon_{K^0}(p)
  &= -\frac{\mst^2}{2\muq} + \sqrt{v^2 p^2 + M_{K^0}^2} ,
\end{eqnarray}
where $v^2=1/3$ at high density.  The CFL-meson masses are given by
\begin{eqnarray}
 M_{\pi^\pm}^2 &= a(\mup + \mdw)\mst + \chi(\mup + \mdw) ,\nonumber\\
 M_{K^\pm}^2 &= a(\mup + \mst)\mdw + \chi(\mup + \mst) ,\nonumber\\
 M_{K^0}^2 &= a(\mdw + \mst)\mup + \chi(\mdw + \mst) .
\label{eq:CFLmeson}
\end{eqnarray}
Here $a=3\Delta^2/(\pi^2 f_\pi^2)$ with
$f_\pi^2=(21-8\ln2)\muq^2/(36\pi^2)$ at high density and $\chi$
parametrizes the contribution of the $\ua$ anomaly which generates
$\langle\bar{\psi}\psi\rangle$ and therefore contributes to the
CFL-meson masses.

In the absence of the $\ua$-breaking term ($\chi=0$), the energies for
$K^+$ and $K^0$ become negative and kaon condensation occurs for
$\mst\gtrsim m^{1/3}\Delta^{2/3}$ with $m$ being either $\mup$ or
$\mdw$.  In particular, the electron contribution to the thermodynamic
potential in the CFL phase favours the $K^0$ condensation (the
CFL-$K^0$ phase)
\cite{Bedaque:2001je,Kaplan:2001qk,Kryjevski:2004cw,Basler:2009vk}.
The phase structure with the CFL-$K^0$ state and its variants have
been also investigated in the NJL-type model
\cite{Forbes:2004ww,Warringa:2006dk}.  The onset of the $K^0$
condensation depends on the strength of the $\ua$ anomaly $\chi$ as is
evident from \eref{eq:CFLmeson}.
 
In view of \eref{eq:CFLmeson} the meson masses have the ordering
$M_{\pi^\pm} > M_{K^\pm}\simeq M_{K^0}$ for $\mst\gg\mup\approx\mdw$
and $\chi\approx 0$, which is inverse of the ordinary ordering in the
vacuum \cite{Son:1999cm}.  This is, however, natural from the diquark
picture as already implied by the order parameter \eref{eq:chiral-CFL}
in which CFL-$\sigma$ meson consists of two diquarks, \ 
$\bar{q}\bar{q}qq$.  The Nambu-Goldstone bosons are accordingly
composed from $\bar{q}\bar{q}qq$;  CFL-$\pi^+$ contains a
$\bar{d}\bar{s}$ diquark that transforms like $u$ quark and an $su$
diquark like $\bar{d}$ quark, while CFL-$K^+$ a $\bar{d}\bar{s}$
diquark and a $ud$ diquark like $\bar{s}$ quark.  Therefore CFL-$K^+$
has a $d$ quark instead of an $s$ quark as compared to CFL-$\pi^+$ and
thus it becomes lighter than CFL-$\pi^+$.  The effect of $\chi\neq0$
in \eref{eq:CFLmeson}, on the other hand, favours the standard
ordering.  This is because $\chi$ arises from the instanton
interaction and induces a mixture of $\bar{q}\bar{q}qq$ and
$\bar{q}q$, which is embodied in the $\Delta^2\sigma$ term in
\eref{eq:3F-Omega}.  If the quark-hadron continuity is realized, not
only the pseudo-scalar mesons but also the vector mesons and fermions
would obey the spectral continuity.  For example, the continuity of
flavour-octet vector mesons at low density and the colour-octet gluons
at high density together with the fate of the flavour-singlet vector
meson have been investigated in the in-medium QCD sum rules
\cite{Hatsuda:2008is}.

In order to clarify the existence of another critical point associated
with the quark-hadron continuity and the formation of $K^0$
condensation in the CFL phase, it is demanded as a theory task to
quantify how much the $\ua$ symmetry is effectively restored at finite
$T$ and/or $\muq$.


\section{Inhomogeneous states}
\label{sec:inhomogeneous}

In the intermediate density regions of the QCD phase diagram, the
ground state may have inhomogeneity with respect to the condensates.
The conventional $\pi^-$ and $\pi^0$ condensations induced by the
$p$-wave pion-nucleon interaction in nuclear matter and in neutron
matter are well-known examples \cite{Takatsuka:1993pv}.  In this
section, we will address some of the proposed inhomogeneous phases
associated with chiral transition and with colour superconductivity.


\subsection{Chiral-density waves}
\label{sec:wave}


\begin{figure}
 \begin{center}
 \includegraphics[width=0.5\textwidth]{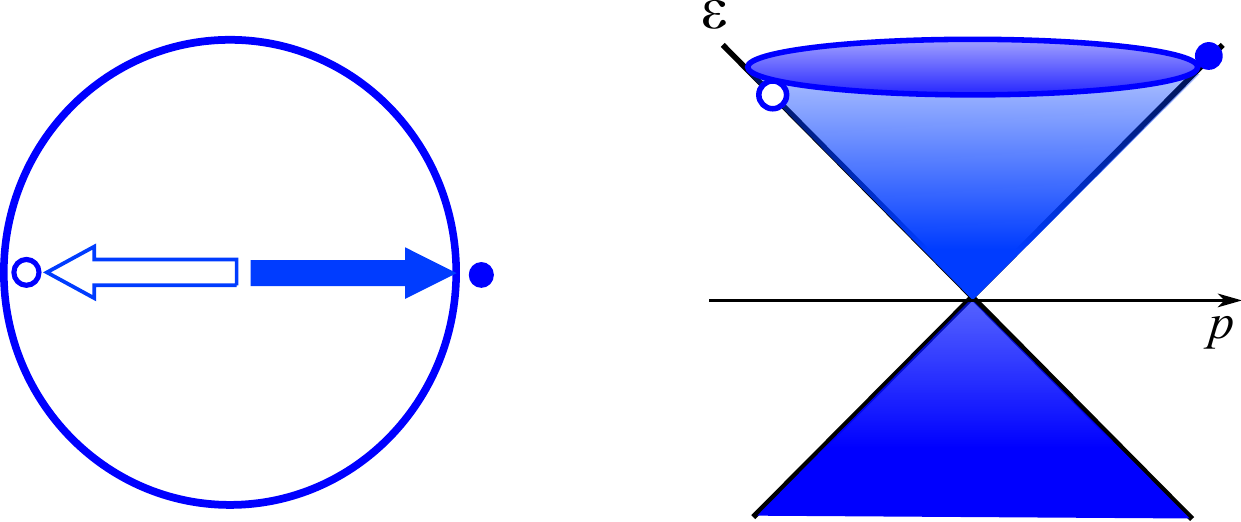}
 \end{center}
 \caption{Particle-hole pair near the Fermi surface generating an
   inhomogeneous chiral condensate with a net momentum, the magnitude
   of which is given by $2\muq$.}
 \label{fig:densitywave}
\end{figure}


In various materials typically at low dimensions, the charge-density
wave (CDW) \cite{Gruner:1988zz} and the spin-density wave (SDW)
\cite{Gruner:1994zz} are realized as ground states.  Their counterpart
in quark matter is the chiral-density waves:  Such a possibility has
been discussed in the large $\Nc$ limit of QCD
\cite{Deryagin:1992rw,Shuster:1999tn} with a spatially modulated
chiral condensate,
\begin{equation}
 \langle\bar{\psi}(x)\psi(x)\rangle = \sigma\cos(2\bq\cdot\bx) ,
\end{equation}
with $|\bq|=\muq$.  \Fref{fig:densitywave} is a schematic picture of
the pairing of a quark and a quark-hole which makes an inhomogeneous
condensate.  The magnitude of the net momentum is given by $2\muq$.
This pairing is favoured by the forward scattering with singular
interaction $\sim 1/q^2$ at high density.  The conclusion of
\cite{Shuster:1999tn} is that this chiral-density wave state can be
realized only for $\Nc\gtrsim 1000\Nf$.  The reason why large $\Nc$ is
required is that the gluon interaction $\sim 1/q^2$ is IR screened
(Thomas-Fermi screening and Landau damping) by the quark loops of
$O(1/\Nc)$.

In \cite{Nakano:2004cd} a different ans\"{a}tz in $2$-flavour case has
been investigated;
\begin{equation}
 \langle\bar{\psi}(x)\psi(x)\rangle
  - \rmi \langle\bar{\psi}(x)\rmi\gamma^5\tau_3 \psi(x)\rangle
  = \sigma \,\rme^{2\rmi \bq\cdot\bx} .
\end{equation}
Here $\tau_3$ is a third component of the flavour Pauli matrices.
This type of condensate is known to be realized as the
\textit{chiral spiral} in $(1+1)$-dimensional chiral models at finite
density \cite{Schon:2000qy}.  Recently the chiral spiral in QCD is
analyzed with confinement effects taken into account
\cite{Kojo:2009ha}.  The confining interaction $\sim 1/(\bq^2)^2$ is
more IR singular than the perturbative interaction, so that the
pairing like drawn in \fref{fig:densitywave} may be realized for
smaller $\Nc$ than discussed in \cite{Shuster:1999tn}.

In addition to the IR singular forward scattering, there is another
source to produce a spatial modulation of the chiral condensate.  To
see this, the Ginzburg-Landau approach is again useful.  The general
terms of the free energy up to the sixth order
 including derivatives are
\cite{Nickel:2009ke}
\begin{equation}
\hspace{-3em}
 \Omega = c_2 M^2 + c_4 M^4 + c_4' (\nabla M)^2
  + c_6 M^6 + c_6' (\nabla M)^2 M^2 + c_6'' (\Delta M)^2 ,
\end{equation}
where $M$ is an order parameter for the chiral phase transition
(either the chiral condensate or the constituent quark mass in the
chiral limit).  Note that the magnitude of $M$ and its spatial
variation are assumed to be the same order.  If $c_4'$ is negative,
inhomogeneous $M$ tends to be favoured.  However, the sign of $c_4'$
cannot be determined by symmetry argument alone.  In the mean-field
treatment of the $2$-flavour NJL model, the expansion turns out to be
\begin{equation}
\hspace{-3em}
 \Omega = \frac{\alpha_2}{2}M^2 + \frac{\alpha_4}{4}
  \bigl[ M^4 + (\nabla M)^2 \bigr]
  + \frac{\alpha_6}{6}\Bigl[ M^6
  + 5(\nabla M)^2 M^2 + \frac{1}{2}(\Delta M)^2 \Bigr] ,
\end{equation}
where $\alpha_2$, $\alpha_4$ and $\alpha_6$ are all calculable within
the model.  In particular, the sign of $\alpha_4$ controls whether the
inhomogeneous state occurs:  If it is negative, the first-order
transition is driven and simultaneously a spatial modulation
develops.  Therefore the first-order phase boundary is naturally
surrounded by the inhomogeneous state in the QCD phase diagram as
illustrated in \fref{fig:phase}.  Essentially the same phenomena takes
place in the chiral Gross-Neveu model in $(1+1)$-dimensions
\cite{Schon:2000qy}.

Importance of inhomogeneous phases in the QCD phase diagram has been
revisited recently and physics implications have not been fully
unveiled.  Especially it is an urgent but unanswered question how the
inhomogeneous condensate could affect the region around, if any, the
QCD critical point.


\subsection{Implications from chromomagnetic instability}
\label{sec:chromomagnetic}

A similar inhomogeneity arises not only in the chiral condensate but
also in the diquark condensate $\Delta(x)$.  The CFL phase is rigid
and $\Dud\simeq\Dds\simeq\Dsu\simeq\Delta$ as long as
$\delta\muq=\mst^2/\muq<2\Delta$.  Once the Fermi surface mismatch
$\delta\muq$ exceeds this bound, a new form of colour
superconductivity appears, which is a counterpart of what is known as
the Sarma state in condensed matter physics.  The Sarma state is,
however, not stable as it is, and it has been argued that a number
constraint such as the charge neutrality condition may help the
stabilization \cite{Liu:2002gi}.  In the QCD context
\cite{Gubankova:2003uj} such superconducting states are called the
gapless 2SC (g2SC) phase \cite{Shovkovy:2003uu,Huang:2003xd} and
gapless CFL (gCFL) phase \cite{Alford:2003fq,Alford:2004hz} in the
$2$-flavour and $3$-flavour cases, respectively.

\begin{figure}
 \begin{center}
 \includegraphics[width=0.6\textwidth]{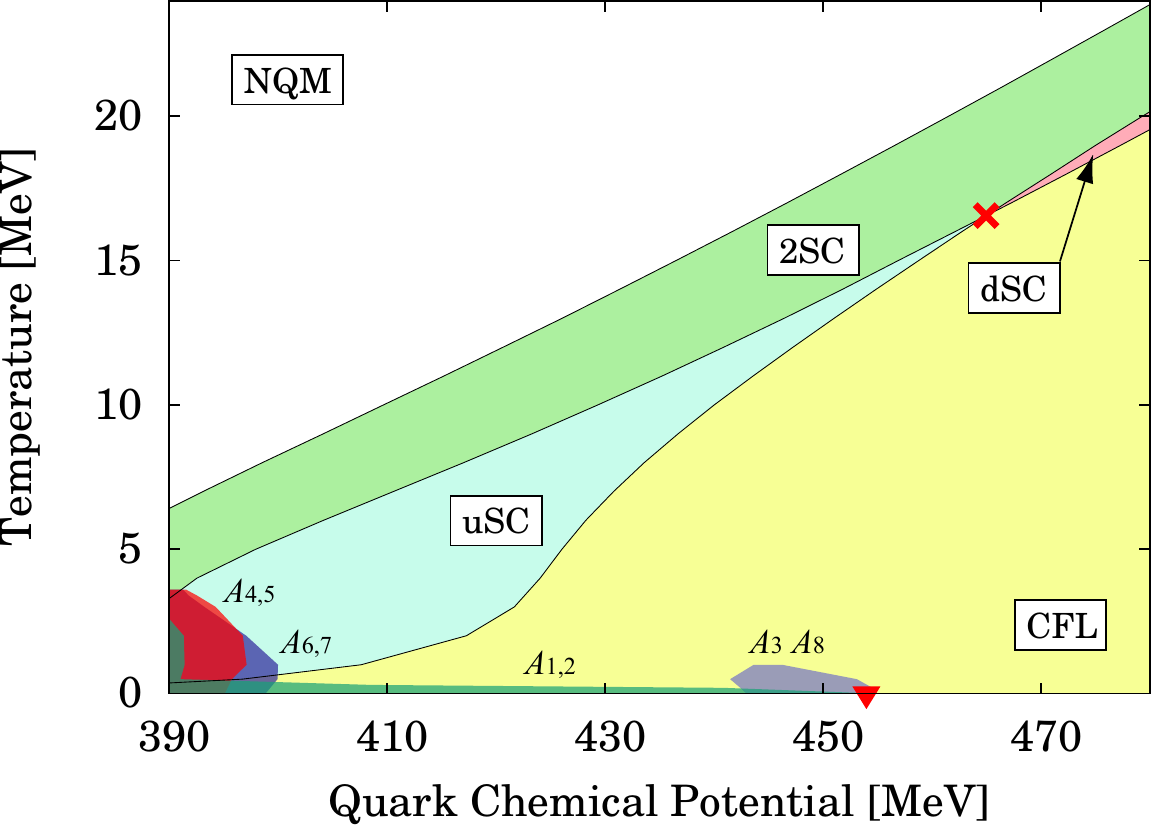}
 \end{center}
 \caption{An example of the mean-field model phase diagram with
   regions suffering from chromomagnetic instability
   \cite{Fukushima:2005fh}.  At high density the ordering is the
   CFL, dSC and 2SC phases from the bottom to the top as explained in
   \fref{fig:fermisphere-gl}.  At lower density the uSC phase is
   favoured because the Fermi surface mismatch between $d$ and
   $s$ quarks is the largest and $\Dds$ tends to melt first when the
   critical $T$ is not large.  The diagram is drawn with sufficiently
   large UV cutoff $\Lambda=1\GeV$ and $\Delta\simeq 40\MeV$ is
   chosen at $\muq=500\MeV$, so that the results are robust against
   cutoff artifacts;  otherwise unphysical structures would easily
   enter.}
 \label{fig:phase-dg}
\end{figure}

It has been found, however, that the gapless superconducting states
suffer from another instability problem.  In
\cite{Huang:2004bg,Huang:2004am,Casalbuoni:2004tb,Alford:2005qw,%
Fukushima:2005cm} the Meissner (magnetic screening) masses in the g2SC
and gCFL phases have been calculated and turned out to be imaginary.
That is, there appear negative eigenvalues from the mass-squared
matrix in colour space,
\begin{equation}
 (m_M^2)^{\alpha\beta} = \frac{1}{3}\sum_{i=1}^3
  \frac{\partial^2\Omega[A]}{\partial A_i^\alpha \partial A_i^\beta}
  \biggr|_{A=0} ,
\end{equation}
for transverse gluons $A_i^\alpha$.  This fact is commonly referred to
as the \textit{chromomagnetic instability}.  The physics implication
of the chromomagnetic instability can be nicely articulated by using
the Ginzburg-Landau description.  A gauged kinetic term can be added
to the free energy in a form of
\begin{equation}
 \Omega[\Delta,A] \;\sim\; \Omega_0[\Delta]
  -\kappa^{ab} [(\partial_i -\rmi gA_i)\Delta]^{\dagger a}
  [(\partial^i-\rmi gA^i)\Delta]^b ,
\end{equation}
from which one can derive the Meissner mass-squared as
$(m_M^2)^{\alpha\beta}\sim
2\kappa^{ab}(T^\alpha)_{ca}(T^\beta)_{bd}\Delta^c\Delta^d$.  We have
$\kappa^{33}<0$ leading to $(m_M^2)^{88}<0$ in the g2SC phase and also
$\kappa^{11}=\kappa^{22}<0$ leading to
$(m_M^2)^{44}$---$(m_M^2)^{77}<0$ whose onset is slightly delayed
after the gapless onset.  The remaining three gluons are unscreened.
On the other hand, in the gCFL phase, all eight eigenvalues of the
mass-squared matrix can potentially become negative
\cite{Casalbuoni:2004tb,Fukushima:2005cm,Fukushima:2005fh}.  The
unstable regions are mapped out in the NJL model
\cite{Fukushima:2005fh} onto the phase diagram, which is presented in
\fref{fig:phase-dg}.  The shaded regions are unstable with respect to
gluons with their colour labelled aside.


Now let us consider a modulation of the diquark condensate in the
simplest form of the plane-wave type.  This is written as
\begin{equation}
 \Delta(x) = |\Delta|\, \rme^{\rmi T^\alpha \bq^\alpha\cdot\bx} ,
\end{equation}
which is known as the (coloured) Fulde-Ferrell-Larkin-Ovchinnikov
(FFLO) state \cite{Giannakis:2004pf,Fukushima:2006su}.  Then, the
stability with respect to growth of $\bq^\alpha\neq0$ is deduced from
the potential curvature;
\begin{equation}
 \frac{1}{3}\sum_{i=3}^3
  \frac{\partial\Omega}{\partial q_i^\alpha \partial q_i^\beta}
  \;\propto\; (m_M^2)^{\alpha\beta} .
\end{equation}
Therefore the chromomagnetic instability leads to the FFLO state.  In
the g2SC case with $\kappa^{33}<0$ a non-coloured ($\alpha=0$) FFLO
state may be enough to cure the chromomagnetic instability
\cite{Gorbar:2005tx} (see also \cite{Giannakis:2005vw} for results in
opposition), but it is a subtle problem if the non-coloured FFLO can
stabilize in all colour channels of gluons
 for $3$-flavour quark matter \cite{Ciminale:2006sm}.

Generally the instability analysis tells us a tendency towards the
destination at best, but cannot go into the ground state.  There are
several candidates proposed so far, among which the free energy should
be compared to sort out the most stable ground state, which is still
an open question.  Below some of proposals are briefly reviewed.


\paragraph{Crystalline colour superconductivity{\rm :}}


\begin{figure}
 \begin{center}
 \includegraphics[width=0.6\textwidth]{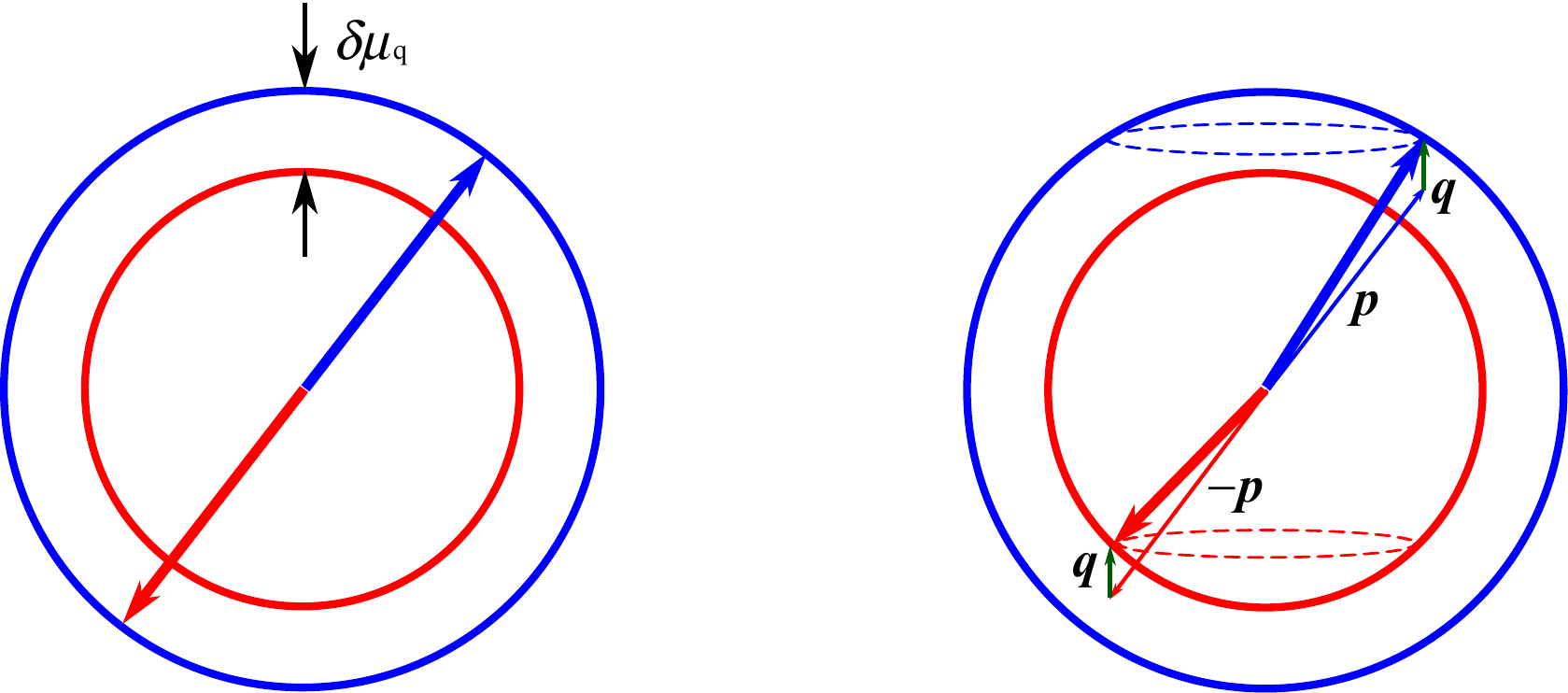}
 \end{center}
 \caption{Pairing with a Fermi surface mismatch by $\delta\muq$.
   Left) BCS pairing with vanishing net momentum.  One particle is
   lifted up from the inner Fermi surface by $\delta\muq$, which costs
   an energy.  Right) FFLO pairing between particles having momenta
   $\bp+\bq$ and $-\bp+\bq$ with a net momentum $2\bq$.  The energy
   costs are reduced around the dotted rings depicted in the
   figure.}
 \label{fig:FFLO}
\end{figure}


It is intuitively understandable that the plane-wave FFLO state is a
likely alternative of the ground state if the Fermi surface mismatch
grows large.  Naturally one can anticipate that $|\bq|$ should be of
order of $\delta\muq$ and actually $|\bq|=1.2\delta\muq$ is found in
the weak-coupling analysis \cite{Alford:2000ze}.  The energy gain
comes from the ring regions on the Fermi surface as indicated in
\fref{fig:FFLO}.  Thus, it is likely that a superposition of multiple
$\bq$'s could further decrease the energy by covering the Fermi
surface with rings.  The general ans\"{a}tz is then,
\begin{equation}
 \langle u(x)d(x)\rangle \sim \Dud\! \sum_{\bq_3^a\in\{\bq_3\}}
  \rme^{2\rmi\bq_3^a\cdot\bx}
\end{equation}
for two flavours and, in addition to this, two more condensates,
\begin{equation}
\hspace{-3em}
 \langle s(x)u(x)\rangle \sim \Dsu\! \sum_{\bq_2^a\in\{\bq_2\}}
  \rme^{2\rmi\bq_2^a\cdot\bx} ,\quad
 \langle d(x)s(x)\rangle \sim \Dds\! \sum_{\bq_1^a\in\{\bq_1\}}
  \rme^{2\rmi\bq_1^a\cdot\bx} ,
\end{equation}
for three flavours \cite{Casalbuoni:2005zp}.  Such a crystal structure
has been investigated by means of the Ginzburg-Landau approach
\cite{Rajagopal:2006ig} with an approximation that $\Dud\simeq\Dsu$
and $\langle d(x)s(x)\rangle\simeq0$ because the Fermi surface
mismatch between $d$, $u$ and $s$ quarks is all the same and the
$d$-$s$ pairing is least favoured (see \fref{fig:fermisphere-gl}).
Then, $\bq_3^a$ and $\bq_2^a$ are optimized so that the free energy
becomes smaller.  In the $2$-flavour case the ground state takes the
face-centred-cubic (FCC) structure with eight $\bq_3^a$ vectors.  In
the $3$-flavour case, a most stable structure is the 2Cube45z that
forms two cubic sets by $\bq_2^a$ and $\bq_3^a$ with a relative
rotation by $45^\circ$ around the $z$ axis.  Another most stable
structure is the CubeX where two rectangles by $\bq_2^a$ and $\bq_3^a$
cross in the X shape to form a cube.  It is desirable to study the
crystallography in microscopic models:  A recent step towards such
direction can be seen e.g.\ in  \cite{Nickel:2008ng}.


\paragraph{Gluonic phase{\rm :}}

The most straightforward interpretation of the chromomagnetic
instability would be the \textit{gluonic phase} \cite{Gorbar:2005rx}
in which gauge fields have a finite expectation value.  If only one
component of gluons condenses, it is equivalent with the single
plane-wave FFLO state.   In fact, an ans\"{a}tz,
\begin{equation}
 \mu_8 = \frac{\sqrt{3}}{2}g\langle A_0^8\rangle , \quad
 \mu_3 = g\langle A_0^3\rangle , \quad
 g\langle A_z^6\rangle \neq0,
\end{equation}
is considered and named the gluonic cylindrical phase in
\cite{Hashimoto:2007ut}.  Because only one spatial gluon condensate is
involved, this state can be mapped to a FFLO-type pairing with a
coloured phase factor.  On the other hand, multi-gluon condensation
cannot be transformed into a single plane-wave FFLO state.  In the
$2$-flavour model the free energy of the preferred gluonic phase has
turned out to be smaller than that of the FFLO state in a wide
parameter region \cite{Kiriyama:2006ui}.  Such an example is the
gluonic colour-spin locked (GCSL) phase that is defined by the
following ans\"{a}tz,
\begin{equation}
 \mu_8 = \frac{\sqrt{3}}{2}g\langle A_0^8\rangle , \quad
 g\langle A_y^4\rangle = g\langle A_z^6\rangle \neq0,
\end{equation}
which is free from the chromomagnetic instability
\cite{Hashimoto:2008wz}.  It is then found that these gluonic phases
have a smaller free energy than the single plane-wave FFLO state and
the unstable g2SC phase.  In most regions the GCSL is the most stable
apart from the vicinity of the first-order phase transition to normal
quark matter where the gluonic cylindrical phase is energetically
favoured.  It has not been understood, however, how to generalize the
description of the gluonic phase to the $3$-flavour case.  Moreover,
an extension of the gluonic condensation with spatial inhomogeneity
may further decrease the free energy \cite{Ferrer:2007uw}.


\paragraph{Meson supercurrent state{\rm :}}

As we have already emphasized, the CFL phase spontaneously breaks
chiral symmetry, where low-energy excitations are described by a
chiral effective Lagrangian expressed in terms of colour-singlet
modes.  Then it is possible to formulate the instability by using the
chiral effective Lagrangian \cite{Schafer:2005ym}.  The origin of the
instability should be common, but it is no longer the
``chromomagnetic'' instability since gluons do not appear explicitly
but all the physical degrees of freedom are Nambu-Goldstone bosons.
Then, instead of gluons, one may expect the condensation of vector
fields given by mesons, that is, the meson currents.  Such a
destination  of the ground state is called the
\textit{meson supercurrent state}.  The description looks different at
a glance from the chromomagnetic instability and the single plane-wave
FFLO state, but the underlying physics must be closely related.  An
advantage of using the chiral effective Lagrangian is that the
inclusion of the $K^0$ condensate is straightforward, which is
complicated in microscopic models.  Such an interpretation as the
supercurrent generation is also proposed in \cite{Huang:2005pv} not
relying on the chiral effective Lagrangian but in terms of phase
fluctuations around the diquark condensate.


\paragraph{Mixed phase and phase separation{\rm :}}

All the states as we have seen so far should compete with a rather
conventional possibility, i.e.\ the mixed phase.  Neutrality
conditions with respect to gauge charge enforce the Fermi surface
mismatch, but  these conditions may be relaxed by a formation of a
mixture of CSC and NQM regions as discussed in the $2$-flavour case in
\cite{Reddy:2004my}.

To clarify the mixed phase structure, however, the precise
determination of the surface tension is indispensable.  The balance
between the surface tension and the Coulomb energy fixes the typical
domain size of the mixed phase structure.  Naturally, the larger the
surface tension is, the larger the favoured domain size grows, and
eventually the mixed phase should be rather regarded as the phase
separation \cite{Iida:2006df}.  It is not easy to extract the
information on reliable value of the surface tension in the
intermediate density region, however.  So far, the possibility of the
mixed phase in $3$-flavour CSC phase has not been studied seriously.


\section{Suggestions from QCD-like theories}
\label{sec:suggestions}

In the research towards the phase diagram of dense QCD, some knowledge
from QCD-like theories would provide us with a useful hint to attack
the QCD problem.  There are many such attempts and it is impossible to
cover all of them in this article.  Here, we discuss some selected
topics which are highly relevant to the understanding of the dense-QCD
phase diagram.


\subsection{Quarkyonic matter at large $\Nc$}
\label{sec:quarkyonic}


\begin{figure}
 \begin{center}
  \includegraphics[width=0.4\textwidth]{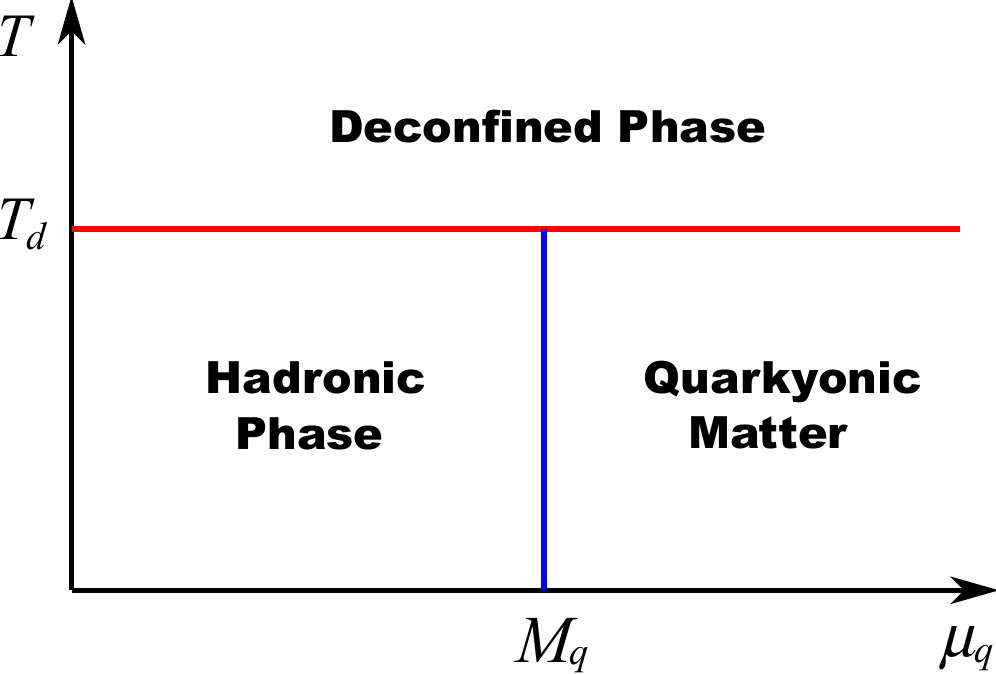}
 \end{center}
 \caption{Schematic phase diagram of large-$\Nc$ QCD.  The pressure is
   $O(\Nc^0)$ in the hadronic phase, $O(\Nc)$ in the quarkyonic
   matter, and $O(\Nc^2)$ in the deconfined phase.}
 \label{fig:quarkyonic}
\end{figure}


The novel QCD phase structure in the large $\Nc$ limit has been
recently proposed \cite{McLerran:2007qj} as schematically shown in
\fref{fig:quarkyonic}.  When $\muq$ is smaller than the threshold of
the constituent quark mass $\Mq \sim \MB/\Nc \sim O(\Nc^0)$, we have
the hadronic phase with zero baryon density at low temperature.  As
the temperature is increased, there appears the first-order
deconfinement transition at $\Td\sim\LQCD$ at which the number of
degrees of freedom and the pressure jump discontinuously from
$O(\Nc^0)$ to $O(\Nc^2)$.  Since quark loops are suppressed by
$1/\Nc$ as compared to gluon contributions, $\Td$ is independent of
$\muq$ in this region as shown in \fref{fig:quarkyonic}.

When $\muq$ becomes greater than $\Mq$, a non-zero baryon density is
turned on.  The pressure associated with this threshold at
$\muq\simeq\Mq$ changes discontinuously from $O(\Nc^0)$ to
$O(\Nc)$.  Also $\Td$ does not change with increasing $\muq$ as long
as $\muq \sim O(\Nc^0)$, so that quarks are still confined in the
right-bottom region ($\muq>\Mq$ and $T<\Td$) in \fref{fig:quarkyonic}.
The confining phase with the pressure of $O(\Nc)$ is called the
\textit{quarkyonic matter} \cite{McLerran:2007qj}.

\begin{figure}
 \begin{center}
 \includegraphics[width=0.24\textwidth]{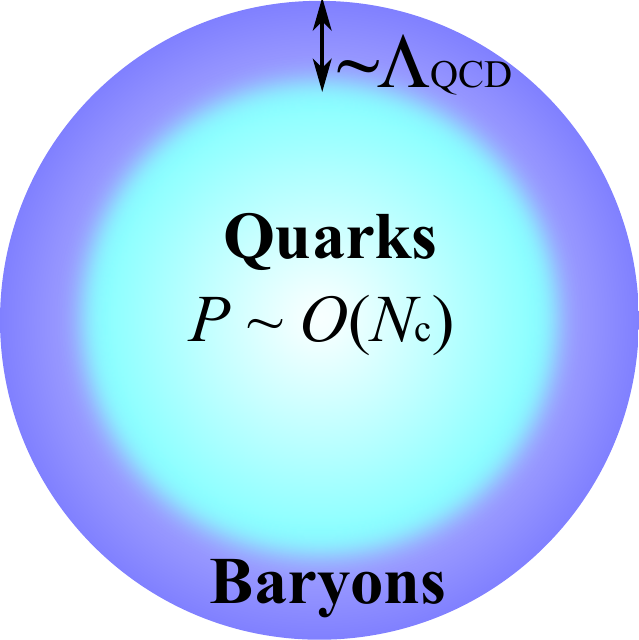}
 \end{center}
 \caption{Intuitive picture of the Fermi sphere which accommodates
   quarkyonic matter ---
    quark Fermi sea and baryonic Fermi surface.}
 \label{fig:fermi}
\end{figure}

If we describe the quarkyonic matter as a weakly interacting quark
system, the pressure of $O(\Nc)$ is a natural consequence, but it is
difficult to reconcile it with the confining feature.  If we describe
the quarkyonic matter as a baryonic system, it must be a strongly
interacting matter where the pressure is dominated by baryon
interactions of $O(\Nc)$ rather than the kinetic pressure of
$O(1/\Nc)$.  A possible idea to unify these two descriptions proposed
in \cite{McLerran:2007qj} is illustrated in \fref{fig:fermi};  the
quarks deep inside the Fermi sphere are weakly interacting, because it
is hard to excite these quarks above the Fermi sea due to Pauli
blocking.  On the other hand, the quarks near the Fermi surface with a
shell-width $\sim\LQCD$ are not affected so much from the Pauli
blocking and can interact strongly through IR singular gluons at
large $\Nc$.  Thus, the bulk thermodynamics such as the pressure,
entropy and so on are dominated by the quarks inside the Fermi sphere,
while the physical excitations on top of the Fermi surface are dominated
by colour-singlet mesons and baryons.


Whether the remnant of quarkyonic matter at large $\Nc$ remains in the
QCD phase diagram at $\Nc=3$ is an open question.  Also, how the
chiral transition takes place inside quarkyonic matter is an important
problem to be studied.  An interesting and plausible possibility comes
from the fact that the gluon propagator may be non-perturbative and IR
singular because of the confining nature.  If this is the case, such
IR singular interaction would induce an inhomogeneous chiral
condensate as elucidated in \sref{sec:wave} or the quarkyonic chiral
spiral \cite{Kojo:2009ha}.


\subsection{QCD at $\Nc=2$}

A QCD-like theory with $\Nc=2$ \cite{Kogut:1999iv,Kogut:2000ek} is
also an interesting limit opposite to the case of quarkyonic matter;
small $\Nc$ instead of large $\Nc$.  Two-colour QCD is free from the
sign problem if the number of degenerate flavours is even, so that
Monte-Carlo simulation at finite baryon density is possible
\cite{Nakamura:1984uz}.  Two-colour QCD shares many non-perturbative
features with $\Nc=3$ QCD such as confinement, chiral symmetry
breaking and superfluidity.

In two-colour QCD, baryons consist of quark pairs $qq$ which are
bosons.  Therefore, nuclear matter composed of fermionic baryons in
$\Nc=3$ is replaced by a superfluid state of bosonic baryons in
$\Nc=2$.  The order parameter of superfluidity is the colour-singlet
baryon condensate $\langle qq\rangle$.  To investigate the phase
structure of two-colour QCD, such baryon condensate has been computed
both in numerical simulations and in analytical strong-coupling
expansion \cite{Kogut:2001na,Nishida:2003uj,Hands:2006ve}.  There is
also a suggestive result that supports the  idea of quarkyonic
matter in $\Nc=2$ from lattice simulations \cite{Hands:2010gd}, 
  which is consistent with the model analysis \cite{Brauner:2009gu}.
The correlation functions and excitation spectra can be studied both
analytically and numerically for $\Nc=2$.  In particular, the
low-energy chiral Lagrangian at finite baryon density as well as
Leutwyler-Smilga type spectral sum rules have been derived
\cite{Kanazawa:2009ks}.  In-medium hadron spectra as a function of
the baryon chemical potential have been also investigated
\cite{Muroya:2002ry,Kogut:2003ju,Lombardo:2008vc}.  As can be seen in
these examples, two-colour QCD continues to be a valuable testing
 ground for studying hot/dense QCD at $\Nc=3$.


\subsection{Ultracold atoms}

High density QCD matter and ultracold atomic systems, although
differing by some twenty orders of magnitude in energy scales, share
analogous physical aspects \cite{Baym:2008me}.  Phenomenological
studies of QCD indicate a strong spin-singlet diquark correlation
inside the nucleon \cite{Selem:2006nd}.  Therefore, it may be a good
starting point to model the transition from $2$-flavour quark matter
at high density to nuclear matter at low density in terms of a
boson-fermion  mixture, in which small-size diquarks are the
bosons, unpaired quarks the fermions and the extended nucleons are
regarded as composite Bose-Fermi particles \cite{Maeda:2009ev}.

Recent advances in atomic physics have made it possible indeed to
realize  a boson-fermion mixture in the laboratory.  In
particular, tuning the atomic interaction via a Feshbach resonance
allows formation of heteronuclear molecules, as recently observed in a
mixture of $^{87}$Rb and $^{40}$K atomic vapours in an optical dipole
trap \cite{TK:exp2}.  An analysis of such non-relativistic mixture of
cold atoms indicate that the BCS-like superfluidity of composite
fermion (N=b+f) with a small gap is a natural consequence of the
strong boson-fermion attraction \cite{Maeda:2009ev}, which may explain
the reason why the fermion gap in nucleon superfluidity can be an
order of magnitude smaller than the gap in colour superconductivity.  A
possible correspondence between cold atoms and QCD is summarized in
\tref{TK:tab:CFA-QCD}.  Fuller understanding, both theoretical and
experimental, of the boson-fermion mixture as well as a mixture of
three species of atomic fermions \cite{TK:Cherng:2007,TK:Rapp:2006rx}
may further reveal properties of high-density QCD.

\begin{table}
 \begin{center}
 \begin{tabular}{|c|c|}
  \hline
  cold atoms   & dense QCD \\
  \hline \hline
  b (bosonic atom) & $D$ (spin-0 diquark) \\ \hline
  f$_{\uparrow, \downarrow}$ (fermionic atom)
   & $q_{\uparrow, \downarrow}$ (unpaired quark) \\ \hline
  N$_{\uparrow, \downarrow}$ (b--f molecule)
   & $\mathcal{N}_{\uparrow, \downarrow}$
   ($D$--$q$ bound state = nucleon) \\ \hline
  b--f attraction & gluonic $D$--$q$ attraction \\ \hline
  b--BEC & 2SC  \\ \hline
  N--BCS & nucleon superfluidity  \\ \hline
 \end{tabular}
 \end{center}
 \caption{Correspondence between the boson-fermion mixture in
   ultracold atoms (such as $^{87}$Rb and $^{40}$K mixture) and the
   diquark-quark mixture in $2$-flavour QCD.}
\label{TK:tab:CFA-QCD}
\end{table}


\section{Summary and concluding remarks}
\label{sec:summary}

In this article we reviewed the current status of theoretical
investigations to explore the QCD phase structure at finite
temperature ($T$) and finite baryon chemical potential ($\muB$).
There are (at least) three fundamental states of matter in QCD:  the
hadronic matter with broken chiral symmetry and quark confinement in
the low-$T$ and low-$\muB$ region, the quark-gluon plasma at high $T$
and the colour superconductivity at low $T$ and high $\muB$.  On top
of these states, some exotic phenomena have been conjectured,
e.g.\ the chiral-density waves, the crystalline colour
superconductivity, the gluonic phase, the quakyonic matter and the
quark-hadron continuity, as covered in this article, and even more
possibilities are still developing.  Lattice QCD simulations, the
Ginzburg-Landau-Wilson approach and effective theories of QCD are
useful theoretical tools to study these phenomena.

From the experimental point of view, the QCD phase transitions at high
$T$ with $\muB/T < 1$ can be studies by using high-energy heavy-ion
collisions at RHIC and LHC.\ \ The QCD phase diagram at relatively low
$T$ with $\muB/T \gtrsim 1$ may also be probed in the future
facilities with lower-energy heavy-ion beams.  Besides, recent
attempts to determine the mass and the radius of neutron stars from
X-ray bursts would be an alternative way to access the equation of
state of dense QCD \cite{Ozel:2010fw,Steiner:2010fz}.  Furthermore,
gravitational waves \cite{Yasutake:2007st} and neutrinos from
supernova explosions \cite{Sumiyoshi:2008kw} or cooling processes in
the neutron star \cite{Carter:2000xf} carry valuable information of
dense QCD matter.  Dense QCD shall continue to be one of the most
fascinating theoretical and experimental topics in particle, nuclear
and astro physics.


\vspace{1cm}

K.~F.\ was supported by Japanese MEXT grant (No.\ 20740134) and also
supported in part by Yukawa International Program for Quark Hadron
Sciences.
T.~H.\ was supported in part by the Grant-in-Aid for Scientific
Research on Innovative Areas (No.\ 2004: 20105003)
and by Japanese MEXT grant (No.\ 22340052).

\section*{References}
\bibliographystyle{utphys}
\bibliography{denseQCD,lattice,colorsuper}

\end{document}